\documentclass[preprint]{aastex63}
\usepackage{float}
\usepackage{stfloats}
\usepackage{subfigure}
\usepackage{amsmath}

\received{June 1, 2019}
\revised{January 10, 2019}
\accepted{\today}
\submitjournal{ApJ}

\shorttitle{3D turbulent CME-flare current sheet }
\shortauthors{Ye et al.}

\begin{document}

\title{Three-dimensional simulation of thermodynamics on confined turbulence in a large-scale CME-flare current sheet}

\correspondingauthor{Jing Ye}
\email{yj@ynao.ac.cn}

\author{Jing Ye}
\affiliation{Yunnan Observatories, Chinese Academy of Sciences, P.O. Box 110, Kunming, Yunnan 650216, People's Republic of China}
\affiliation{Yunnan Key Laboratory of Solar Physics and Space Science, P.O. Box 110, Kunming, Yunnan 650216, People's Republic of China}
\affiliation{Yunnan Province China-Malaysia HF-VHF Advanced Radio Astronomy Technology International Joint Laboratory, Kunming, Yunnan 650216, People's Republic of China}

\author{John C. Raymond}
\affiliation{Smithsonian Astrophysical Observatory, 60 Garden Street, MS 15, Cambridge, MA 02138, USA}

\author{Zhixing Mei}
\affiliation{Yunnan Observatories, Chinese Academy of Sciences, P.O. Box 110, Kunming, Yunnan 650216, People's Republic of China}
\affiliation{Yunnan Key Laboratory of Solar Physics and Space Science, P.O. Box 110, Kunming, Yunnan 650216, People's Republic of China}

\author{Qiangwei Cai}
\affiliation{Institute of Space Physics, Luoyang Normal University, Luoyang, Henan 471934, People's Republic of China}
\affiliation{Yunnan Key Laboratory of Solar Physics and Space Science, P.O. Box 110, Kunming, Yunnan 650216, People's Republic of China}

\author{Yuhao Chen}
\affiliation{Yunnan Observatories, Chinese Academy of Sciences, P.O. Box 110, Kunming, Yunnan 650216, People's Republic of China}
\affiliation{University of Chinese Academy of Sciences, Beijing 100049, People's Republic of China}

\author{Yan Li}
\affiliation{Yunnan Observatories, Chinese Academy of Sciences, P.O. Box 110, Kunming, Yunnan 650216, People's Republic of China}
\affiliation{Yunnan Key Laboratory of Solar Physics and Space Science, P.O. Box 110, Kunming, Yunnan 650216, People's Republic of China}

\author{Jun Lin}
\affiliation{Yunnan Observatories, Chinese Academy of Sciences, P.O. Box 110, Kunming, Yunnan 650216, People's Republic of China}
\affiliation{Yunnan Key Laboratory of Solar Physics and Space Science, P.O. Box 110, Kunming, Yunnan 650216, People's Republic of China}
\affiliation{University of Chinese Academy of Sciences, Beijing 100049, People's Republic of China}



\begin{abstract}
Turbulence plays a key role for forming the complex geometry of the large-scale current sheet (CS) and fast energy release in a solar eruption. In this paper, we present full 3D high-resolution simulations for the process of a moderate Coronal Mass Ejection (CME) and the thermodynamical evolution of the highly confined CS. Copious elongated blobs are generated due to tearing and plasmoid instabilities giving rise to a higher reconnection rate and undergo the splitting, merging and kinking processes in a more complex way in 3D. A detailed thermodynamical analysis shows that the CS is mainly heated by adiabatic and numerical viscous terms, and thermal conduction is the dominant factor that balances the energy inside the CS. Accordingly, the temperature of the CS reaches to a maximum of about 20 MK and the range of temperatures is relatively narrow. From the face-on view in the synthetic Atmospheric Imaging Assembly 131 $\mathring{A}$, the downflowing structures with similar morphology to supra-arcade downflows are mainly located between the post-flare loops and loop-top, while moving blobs can extend spikes higher above the loop-top. The downward-moving plasmoids can keep the twisted magnetic field configuration until the annihilation at the flare loop-top, indicating that plasmoid reconnection dominates in the lower CS. Meanwhile, the upward-moving ones turn into turbulent structures before arriving at the bottom of the CME, implying that turbulent reconnection dominates in the upper CS. The spatial distributions of the turbulent energy and anisotropy are addressed, which show a significant variation in the spectra with height.  
\end{abstract}



\section{Introduction} \label{intro}
Coronal Mass Ejections (CMEs) are the most energetic events in the solar system, and they cause major changes of the large-scale magnetic structures in the corona. Various empirical and analytical models are proposed for the initiation and evolution of CMEs, such as Catastrophe model \citep{Lin2000,Lin2001}, Breakout model \citep{Antiochos1999,MacNeice2004},\cite{Titov1999} model, Photospheric converging model \citep{Amari2003,Zhao2017}, Flux emergence model \citep{Archontis2004}, and Tether-cutting model \citep{Moore2001,Jiang2021}. In the process of the solar eruption, the closed magnetic field is highly stretched by the loss of equilibrium and a plate-like reconnection region, namely current sheet (CS), forms between the two magnetic fields of opposite polarity, and magnetic reconnection (MR) occurs in the CS \citep{Mikic1994,Forbes2000,Lin2002b}. Therefore, a long CS is always expected to appear in the major eruption. With one end anchored on the rising flux rope (FR), the CS forms a low pressure area where both magnetic field and plasma from the background are pushed into the CS, as reconnection inflows. In other words, the reconnection process in the plate-like CS is externally driven by the rising FR and the reconnection inflows in the third dimension.\par
 A sufficiently fast reconnection rate is required to support a successful solar eruption \citep{Lin2002a}. The mechanism responsible for driving the fast magnetic reconnection in the CS is an ongoing research topic. The Sweet-Parker model only describes a slow reconnection with a long and thin diffusion region \citep{Sweet1958,Parker1957}, while Petschek model is often used to explain the fast magnetic reconnection process by introducing a single X-type reconnection region \citep{Petschek1964}. Furthermore, plasmoid instabilities \citep{Loureiro2007,Bhattacharjee2009} and turbulence \citep{Strauss1988,Lazarian2012,Lazarian2020} play key roles in efficient diffusion in the reconnecting CS. The linear theory for the tearing mode \citep{Furth1963} suggests that the CS becomes unstable to tearing mode as the aspect ratio exceeds $2\pi$. And \cite{Shen2011} suggest that plasmoid instabilities take place in the solar flares only if the Lundquist number exceeds a critical value of $S\approx 10^4$. The non-linear instability tears the global CS into many small-scale pieces with many plasmoids transferring the magnetic energy from large to small scales.  At the same time, merging magnetic islands create an inverse cascade \citep{Fermo2010}.  The cascades result in power-law spectra in the flux and size distributions with indices near -2 \citep{Barta2011, Shen2013}. The well-developed turbulence in the CS is mainly generated by plasmoid instability in 2D, and the global CS is filled with the diffusion region due to the appearance of plasmoids \citep{Ye2019,Xie2022}. However, 3D and 2D turbulence are fundamentally different in the process of MR \citep{Eyink2011}. The theoretical work of \cite{Lazarian1999} predicts a fast reconnection speed in the presence of weak turbulence, which is numerically verified in 3D by \cite{Kowal2009}. And the turbulent meandering of magnetic field lines has been shown numerically to induce a strong violation of magnetic flux freezing \citep{Eyink2013}. It is shown by \cite{Kowal2017,Kowal2020,Beresnyak2017} that 3D turbulence develops spontaneously in the reconnection layer, and it is not a collection of 2D magnetic loops. \cite{Lazarian2020} also suggest that plasmiod reconnection dominates in 2D, while turbulent reconnection is more prominent in 3D. On the other hand, \cite{Nishida2013} made an effort to extend the 2D model of solar flares to a 3D model, and conclude that the plasmoid-induced reconnection theory is still valid in 3D cases. \cite{Dong2022} recently performed 3D simulations of MHD turbulence at a large magnetic Reynolds number of $R_m=10^6$, and they provide direct evidence that plasmoids can play a role for forming the turbulent energy spectra with an index of -2.2. Thus the mechanism of magnetic reconnection in the realistic turbulent CS is still poorly understood. \par

Many observations suggest that fragmented and turbulent structures exist to enable the reconnection in the CME-flare CS \citep{Lin2007,LiuR2013,Lin2015,Li2018,Cheng2018,Lee2020}. \cite{Patel2020} further accomplish a statistical study of plasmoids in the CS of the CME eruption on September 10, 2017 using extreme-ultraviolet (EUV) and white-light coronagraph images. For the same event, \cite{Warren2018} observe the formation and evolution of the hot CS, consistent with the 2D numerical work of \cite{Ye2020}. It has been also observed by \cite{Hanneman2014,Innes2014,Cai2019} that a distributed plasma structure of about 10 MK, namely supra-arcade fan (SAF), exists above the post-flare loops. And the motion of supra-arcade downflows (SADs) contributes significantly to the plasma heating of the SAF \citep{Reeves2017,Samanta2019}. The formation of the SAF and the SAD were recently shown to be related to the turbulent region above the flare loops by simulations \citep{Cai2021,Shen2022,Ruan2023}. Additionally, \cite{Mei2017} compared the 3D CS formed during the FR eruption with that in 2D models, which are very similar in the planar cut view, but they only considered an isothermal solar atmosphere. \cite{Reeves2019} investigate the plasma heating in the CS using 3D simulations considering thermal conduction and radiative cooling mechanisms, but the eruption is not very energetic, with maximum plasma temperatures about only $\sim 3-5$ MK. However, the direct numerical study on thermodynamics of the large-scale CS and plasmoids in 3D CMEs are still limited. \par
In this paper, we perform full 3D MHD simulations based on \cite{Titov1999} model to study the CME-driven reconnection process. Adaptive mesh refinement (AMR) is utilized to achieve the high resolution grid around the 3D CS, in order to capture the evolution of fine structure within. In section \ref{setup}, the analytical solution for the initial configuration and numerical setups are described. Section \ref{results} shows the numerical results of the global and local evolution. Lastly, we give the discussion and conclusions in Section \ref{conclusion}.

\section{Numerical Description} \label{setup}
The governing MHD equations used in our simulations are described in Cartesian geometry as follows:
\begin{eqnarray}
&&\frac{\partial \rho}{\partial t} + \nabla\cdot(\rho\textbf{v})=0\label{mhd1}\\
&&\frac{\partial e}{\partial t}+\nabla\cdot\left[(e+P^* )\textbf{v}-(\textbf{v}\cdot \textbf{B})\textbf{B}\right]\nonumber\\
&&=\rho\textbf{g}\cdot\textbf{v}+\nabla\cdot\left[ \eta\textbf{B}\times (\nabla\times\textbf{B})-\textbf{F}_C\right]\label{mhd2}\\
&&\frac{\partial(\rho \textbf{v})}{\partial t}+\nabla\cdot\left[\rho\textbf{v}\textbf{v}-\textbf{BB}+P^*\textbf{I}\right]=\rho\textbf{g}\label{mhd3}\\
&&\frac{\partial\textbf{B}}{\partial t}=\nabla\times(\textbf{v}\times \textbf{B}-\eta\nabla\times\textbf{B})\label{mhd4}\\
&&P^* =P+\frac{\textbf{B}\cdot \textbf{B}}{2} \label{mhd5}\\
&&P=\rho T\label{mhd6}
\end{eqnarray}
where $\rho$ is the plasma mass density,  $\textbf{v}$ is the flow velocity, $P$ is the thermal pressure, $\eta$ is the magnetic resistivity, $\textbf{B}$ and $\hat{\textbf{B}}$ are the magnetic field and the associated unit vector, $\gamma=5/3$ is the ratio of specific heats, $\textbf{g}$ is the gravitational acceleration, and $e=P/(\gamma-1)+\rho\textbf{v}^2/2+\textbf{B}\cdot\textbf{B}/2$ is the total energy density. The Spitzer model for thermal conduction \citep{Spitzer1962} is included in Eq.(\ref{mhd2}) in the form of $\textbf{F}_C=-\kappa_{||}(\nabla T\cdot \hat{\textbf{B}})\hat{\textbf{B}}-\kappa_{\perp}(\nabla T-(\nabla T\cdot \hat{\textbf{B}})\hat{\textbf{B}})$. The parallel and perpendicular coefficients $\kappa_{||}, \kappa_{\perp}$ are the same as given in \cite{Ye2021}. Particularly, thermal conduction must saturate as the temperature gradient becomes too large, and the heat flux is limited to the saturated flux of \citet{Cowie1977} (See Appendix). All variables presented here are dimensionless. The normalization units are $L_0=5\times 10^9$ cm, $n_0=10^{10} \textrm{ cm}^{-3}$ and $T_0=10^6$ K
for length, number density and temperature, respectively. Thus, we have the characteristic values $\rho_0=1.673\times 10^{-14}\textrm{g cm}^{-3}$, $B_0=5.89$ G, $P_0=2.76 \textrm{ g cm}^{-1}\textrm{s}^{-2}$, $v_A=128.5\textrm{ km s}^{-1}$ and $t_0=389.15$ s for mass density, magnetic strength, gas pressure, velocity and time. To complete the equations, the divergence-free condition ($\nabla\cdot \textbf{B}=0$) needs to be satisfied at any evolution time. \par
The gravitational acceleration deceases with the height along the z-axis, given in the dimensionless form by
$\textbf{g}=-\rho_0L_0/P_0g_0\hat{\textbf{z}}/(1+zL_0/R_\odot)^2$. Here, we have the gravitational acceleration constant near the solar surface $g_0=2.74\times 10^4\textrm{ cm s}^{-2}$, the solar radius $R_\odot=6.961\times 10^{10}$ cm, and the unit vector $\hat{\textbf{z}}$ in z-direction. In order to hold naturally the line-tied condition at the bottom of the simulation box, we adopted a two-layer gravitationally stratified atmosphere including the chromosphere and the corona. The initial temperature distribution is given simply by
\begin{equation}
T(z) = \left\{ 
\begin{aligned}
&T_{chr},  &&\textrm{for} &0\leq z \leq h_{chr} ,\\
&T_{cor},  &&\textrm{for} & z>  h_{chr},
\end{aligned}
\right.
\end{equation}
where $T_{chr}=0.01$ is the chromosphere temperature, $T_{cor}=1$ is the corona temperature, and $h_{chr}=0.4$ is the top of the chromosphere. We note that the initial temperature distribution is not continuous at the low altitude. The thermal conduction could change the gravitationally stratified layer once the simulations start. Since the associated conduction time at the low layer is much longer than the simulation time, the plasma evolution in higher coronal regions won't be significantly affected in the later phases. Then we have the associated gas pressure distribution due to hydrostatic equilibrium: 
\begin{equation}\label{InitialP}
P(z)=\left\{
\begin{aligned}
&p_{chr}\exp\left[\frac{1}{\lambda T_{chr}}\frac{-z}{1+zL_0/R_\odot}\right], &&\textrm{for} & 0\leq
z \leq h_{chr}, \\
&p_{cor}\exp\left[\frac{1}{\lambda T_{cor}}\frac{(-z+h_{chr})}{(1+zL_0/R_\odot)(1+h_{chr}L_0/R_\odot)} \right], &&\textrm{for} &z> h_{chr} ,
\end{aligned}
\right.
\end{equation} 
with 
\begin{equation}
p_{chr} = p_{cor} \exp\left[ \frac{1}{\lambda T_{chr}}\frac{h_{chr}}{1+h_{chr}L_0/R_\odot} \right], \quad \lambda = \frac{2 k_BT_0}{m_Hg_0L_0}.
\end{equation}
Here, $k_B=1.38\times 10^{-23} \textrm{erg K}^{-1}$ is the Boltzmann constant, $m_H=1.67\times 10^{-24}$ g is the mass of the hydrogen atom, and $p_{cor}$ is the gas pressure set at the height $z=h_{chr}$. 
\par
The initial magnetic configuration is similar to the T \& D model \citep{Titov1999,Torok2004}, which consists of three components. The first component, denoted by $B_I$, comes from a pre-existing flux rope of major radius $R$, carrying a uniformly distributed ring current $I$ of a minor radius $a$. The second one, denoted by $B_q$, is created by a pair of magnetic sources with the distance $L$ at the depth $z=-d_1$ under the photosphere ($z=0$). The two sources of charge $\pm q_1$ are located symmetrically to the axis of the ring to mimic the active region. The detailed formulae for $B_I$ and $B_q$ can refer to \cite{Titov1999}. The third one, denoted by $B_t$, represents a dipole source of strength $q_2$ buried at $z=-d_2$ at the origin of xy-plane, in order to allow the magnetic strength to decay with the square of the height. Following the work by \cite{Mei2020}, the associated magnetic potential vector reads as
\begin{equation}\label{Dipole}
\textbf{A}_t =  \frac{q_2}{(x^2+y^2+(z+d_2)^2)^{3/2}}\left( 
\begin{array}{c}
z+d_2  \\
0 \\
-x
\end{array}
\right)
\end{equation}
On the assumption of a thin enough flux rope (i.e. $a<<R$), we can treat the global and internal equilibrium almost independently \citep{Isenberg1993} .  The global equilibrium in an axisymmetric system is reduced to the balance between the Lorentz Forces caused by the curvature of the tube axis and interaction of the background magnetic field $B_q$ at the given strength $\pm q_0$ with the ring current $I$, which gives the equilibrium current
\begin{equation}
I = \frac{8\pi q_0LR(R^2+L^2)^{-3/2}}{\ln (8R/a)-5/4}.
\end{equation}
Since this work mainly focuses on the detailed evolution after the major eruption, we start the simulation from a non-equilibrium state by setting $q_1=0.7q_0$, in order to have an immediate eruption. \par
On the other hand, the internal equilibrium of the flux rope is realized by the balance between the inward force from the poloidal magnetic field and the outward force from the thermal pressure and magnetic pressure from the toroidal field \citep{Mei2018}. Hence, we have the explicit expression for the gas pressure within the flux rope:
\begin{equation}
P=P_{m}+(1-\alpha)\frac{I^2}{4\pi^2a^2}\left(1-\frac{r_a^2}{a^2}\right), \quad \textrm{for} \quad r_a\leq a,
\end{equation}
with 
\begin{equation}
\begin{aligned}
r_a=\sqrt{x^2+(r_{\perp}-R)^2},\\
r_{\perp} = \sqrt{y^2+(z-d_1)^2}.
\end{aligned}
\end{equation}
Note that $\alpha$ is the ratio of the thermal pressure to magnetic pressure, and $P_m$ denotes the background pressure obtained from Eq.(\ref{InitialP}). Reusing the symbol of the line current $I_0$ in \cite{Titov1999}, the internal toroidal field can be written as 
\begin{equation}\label{Toroidal}
B_t = \frac{I_0}{2\pi}\left[\sqrt{\frac{2\alpha I^2}{a^2 I_0^2} \left( 1-\frac{r_a^2}{a^2} \right)+\frac{1}{R^2}}+\sqrt{y^2+(z+d_1)^2}-\frac{1}{R}\right], \quad r_a\leq a.
\end{equation}
In this way, we only need to adjust the value of $I_0$ to make the external field derived from Eq.(\ref{Dipole}) and the internal field of Eq.(\ref{Toroidal}) the same at the surface of the flux rope \citep{Mei2018}. In addition, the flux rope is supposed to carry the cold plasma from the chromosphere \citep{Mackay2009}, so the temperature within is set to $T = 0.02$ for $r_a\leq a$. Therefore, the initial density distribution is eventually obtained by Eq.(\ref{mhd6}).\par
The simulation box is defined in 3D as $(x,y,z)\in [-4,4]\times[-4,4]\times[0,12]$, and the root grid size is $80\times 80\times 120$. For the study of 3D small-scale structures in the CS, it is necessary to carry out the appropriate mesh refinement techniques to save computational sources. Practically, the static mesh refinement (SMR) of maximally 3 levels is utilized before the computation starts: level 1 for $(x,y,z)\in [-0.8,0.8]\times [-1.6,1.6]\times [0,5]$, level 2 for $(x,y,z)\in [-0.4,0.4]\times [-0.8, 0.8]\times [0,4.5]$, level 3 for $(x,y,z)\in [-0.2,0.2]\times[-0.6,0.6]\times [0.5,4]$. And the adaptive mesh refinement (AMR) of 5 levels is activated in the region of $[-0.4,0.4]\times [-0.5,0.5]\times[0.5,5]$ around the CS for $t\geq 0.4$ and the finest cell size corresponds to a physical length of $156.25$ km. The essential parameters for our model are given as $L=0.4$, $R=1.2$, $a=0.2$, $d_1=0.1$, $d_2=1.2$, $q_0=40.75$, $q_2=2.92$ and $\alpha = 0.997$. The gas pressure at $z=h_{chr}$ is $p_{cor}=1$, which gives the plasma $\beta\approx 0.15$ at the bottom of the corona. The normalized resistivity is set as $\eta=10^{-5}$, resulting in a diffusion coefficient $\eta_c=\eta L_0v_A=6.425\times 10^{11} \textrm{ cm}^2\textrm{s}^{-1}$ in cgs units. \par
The boundary conditions included are similar to the 2D case of \cite{Ye2021}. That is the line-tied condition at the bottom boundary $z=0$ formalized by \cite{Shen2018}, while the open conditions are used for the other 5 boundaries allowing the plasma to enter or exit freely the simulation box. The full 3D simulations are performed by the MPI-parallelized AMR code
``MPI-AMRVAC'' \citep{Keppens2012,Xia2018} using the Godunov style finite volume method in general. Additionally, we choose the Harten-Lax-van Leer approximate Riemann solver \citep{Harten1983} together with a third-order asymmetric flux limiter \citep{Koren1993} to capture the shock structures and a three-step Runge-Kutta time solver. The anisotropic heat flux is solved by the SuperTimeStepping method \citep{Meyer2012}.
We have completed two runs with and without thermal conduction as listed in table \ref{para}. The Appendix shows that the temperature distribution for Run A is much overestimated (larger than 100 MK) leading to the very low density inside the CS, while the distributions for Run B are more consistent with observations. Conduction is a dominant factor that balances the energy inside the CS. It also turns out, from our simulations, that Run B shows the large-scale CS with more flux tube structures appearing than Run A.  In order to better study the turbulent features of the CS, the detailed analyses in the next section are all based on the results from Run B. \par
 \begin{table}[H]
 \caption{Simulation parameters for Run A and B.  \label{para}}
 \centering
 \begin{tabular}{c|c|c|c}
 \hline
 &root grid& AMR levels & Thermal conduction\\
\hline
 Run A& $80\times 80\times120$&5& No\\
 Run B& $80\times 80\times120$&5& Yes\\
 \hline
 \end{tabular} 
 \end{table}
To compare with observations, the synthetic extreme ultraviolet (EUV) emissions are produced by the forward modeling code FoMo (v3.4, \cite{VanDoorsselaere2016}), taking into account contribution functions from the CHIANTI atomic database \citep{DelZanna2015}. The spatial resolution of the EUV images is the same as the observations, with a pixel corresponding to 435 km at 131 $\mathring{A}$ passband.

\section{Simulation Results} \label{results}
\subsection{Global evolution}
\begin{figure}[H]
\begin{interactive}{animation}{global/CSfilm.avi}
\centering
\subfigure[$t=0$]{\includegraphics[width=0.3\textwidth,height=6cm,draft=false]{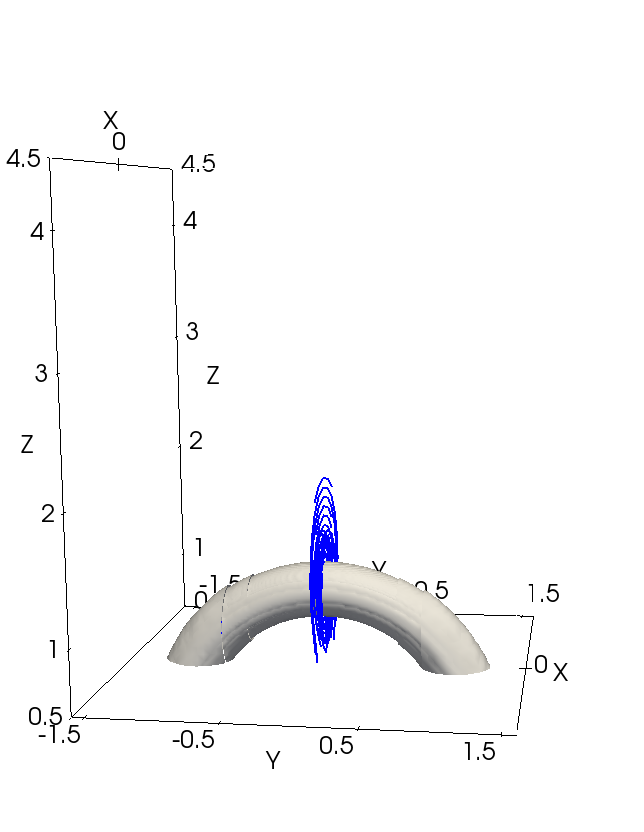}}
\subfigure[$t=0.52$]{\includegraphics[width=0.45\textwidth,height=6cm,draft=false]{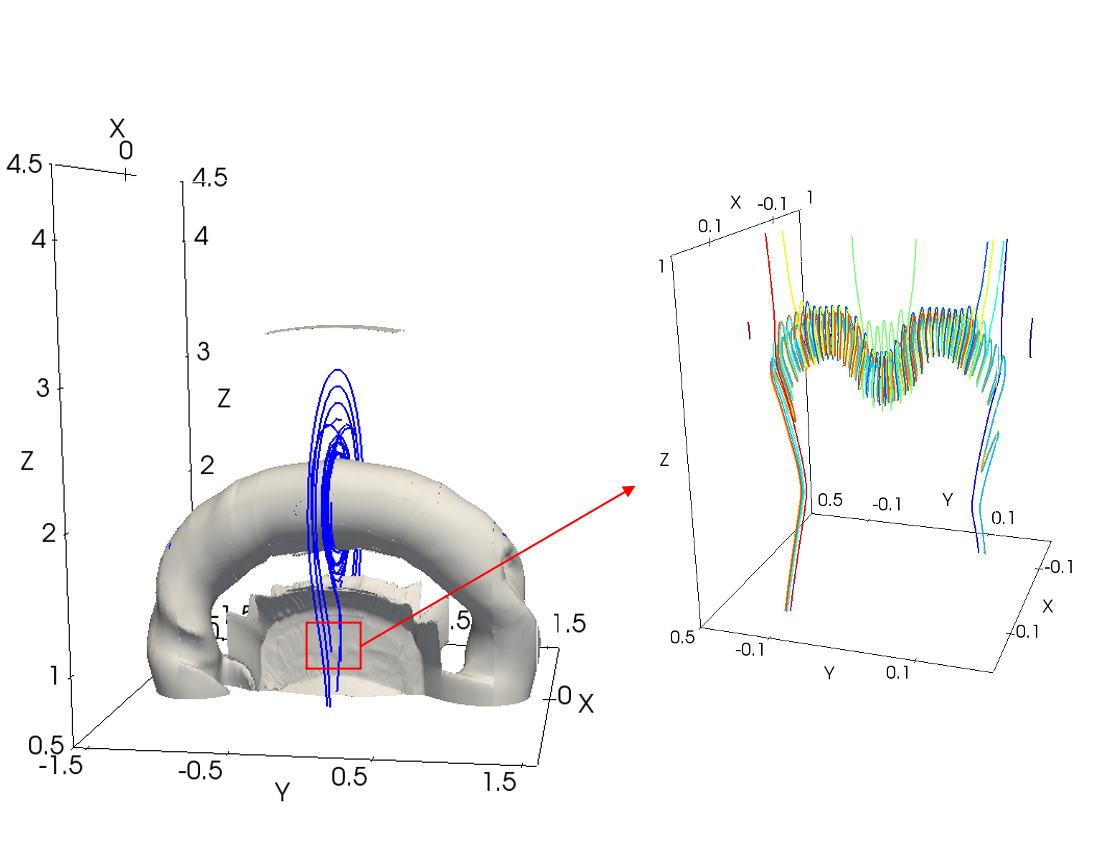}}
\subfigure[$t=0.96$]{\includegraphics[width=0.3\textwidth,height=6cm,draft=false]{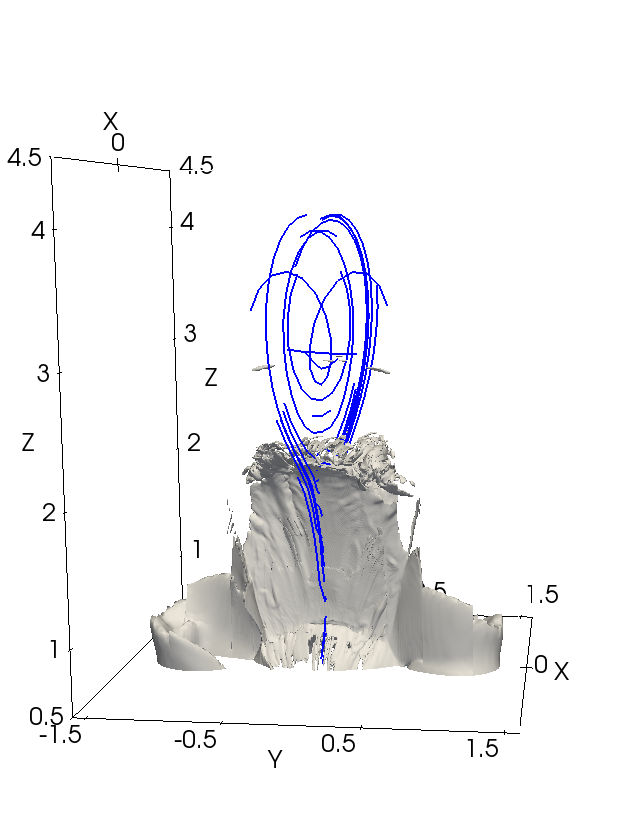}}
\subfigure[$t=1.42$]{\includegraphics[width=0.3\textwidth,height=6cm,draft=false]{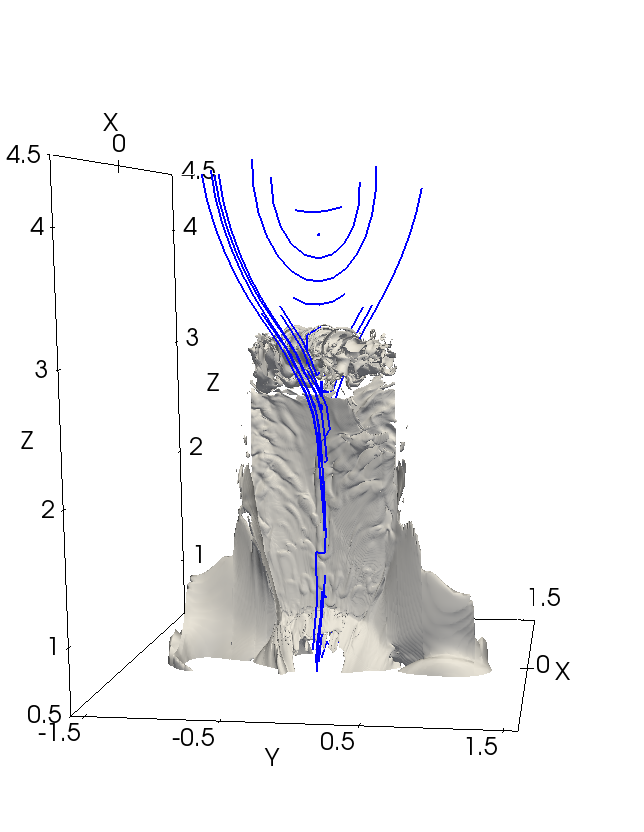}}
\subfigure[$t=2.0$]{\includegraphics[width=0.3\textwidth,height=6cm,draft=false]{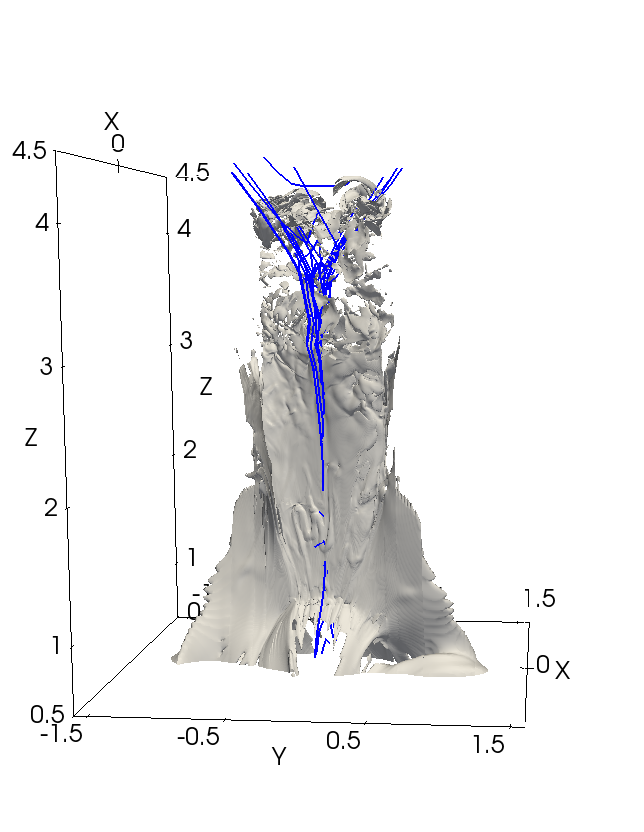}}
\end{interactive}
\caption{\small Snapshots of the global evolution of current density and magnetic fields at different times. The structures of the current sheet are shown by the current isosurface with value $|J|=50$ in white. The blue solid lines are magnetic fields crossing the z-axis. The colors are used in the right panel in (b) for the better visualization of different magnetic fields for the first plasmoid marked in the red box in the left panel. An animation of the detailed evolution inside the 3D CS is available for the entire simulation time. \label{global}}
\end{figure}

\begin{figure}[H]
\centering
\subfigure[]{\includegraphics[width=0.4\textwidth,height=6cm,draft=false]{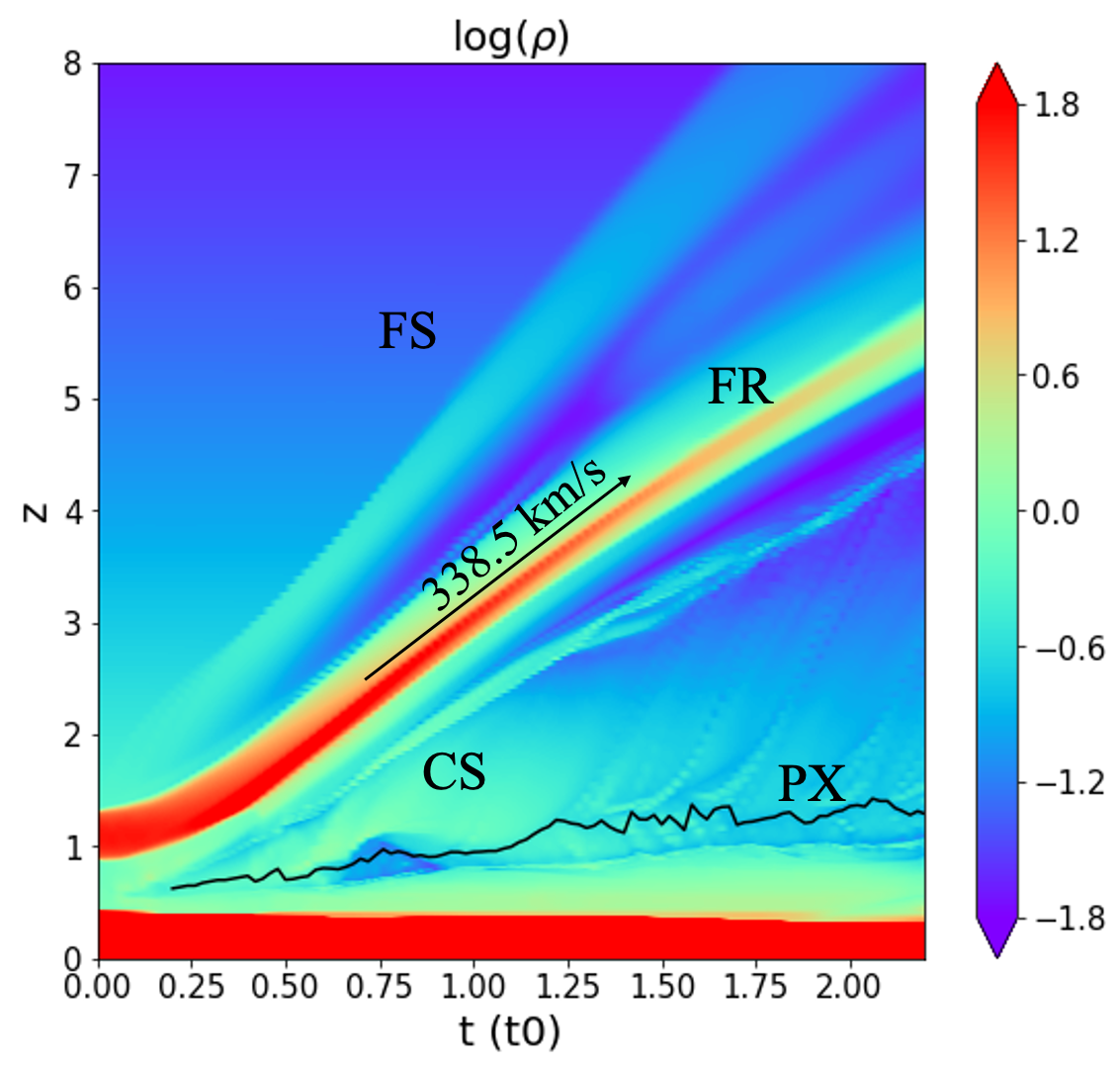}}
\subfigure[]{\includegraphics[width=0.4\textwidth,height=6cm,draft=false]{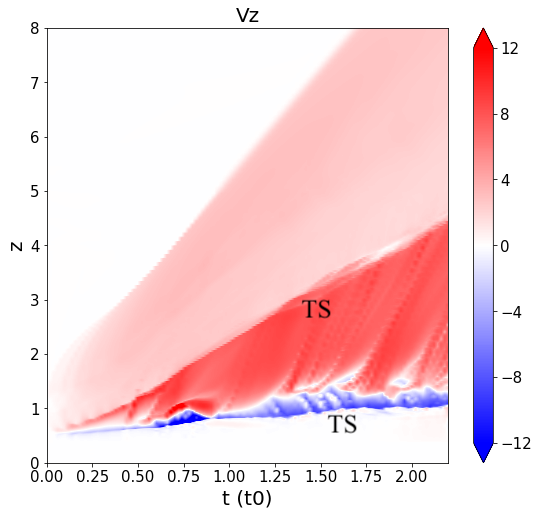}}
\subfigure[]{\includegraphics[width=0.4\textwidth,height=6cm,draft=false]{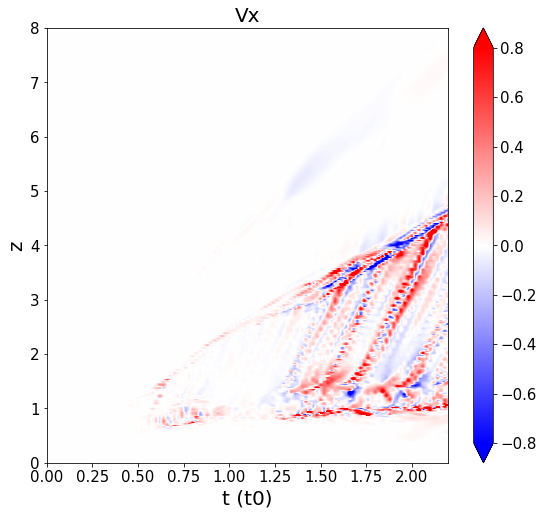}}
\subfigure[]{\includegraphics[width=0.4\textwidth,height=6cm,draft=false]{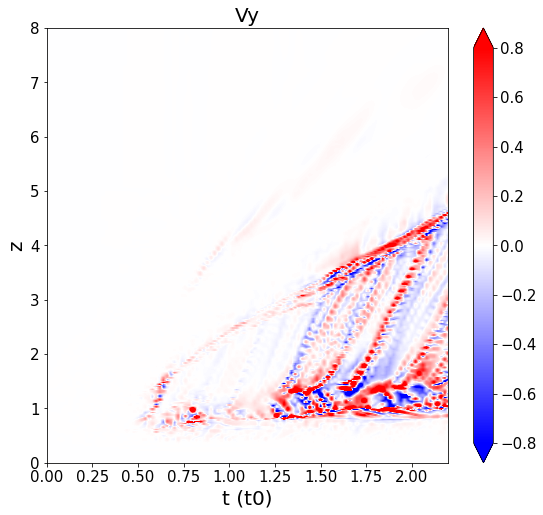}}
\caption{\small Time evolution of the CS along the z-axis passing the origin. (a) Mass density in log-scale; (b) velocity in the $z$-direction; (c) velocity in the $x$-direction; (d) velocity in the $y$-direction. The solid black line in (a) stands for the PX line.\label{tmap} }
\end{figure}

Figure \ref{global} shows the formation and multi-scale processes of the CS by the current isosurface $|J|=50$ following the erupted flux rope (FR) in 3D configurations. A close-up look for the detailed dynamical evolution in the CS is presented in the corresponding animation from $t=0$ to $t=2.2$. We roughly calculate the effective resistivity $\eta_e=V_{\textrm{in}}\delta$ to be about at $ 2.38\times 10^{14}\textrm{ cm}^2\textrm{s}^{-1}$  at $t=1.42$, where $\delta\approx 4.7\times10^7$ cm is the minimum full width of the CS, and $V_{\textrm{in}}\approx 50.7$ km/s is the reconnection inflow velocity near the middle of the CS. Then the resulting Lundquist number is estimated by $S=L_cV_c/\eta_e\approx 1.5\times 10^4$, with $L_c\approx 2RL_0=1.2\times 10^{10}$ cm the characteristic FR length and $V_c\approx 2904.1$ km/s the Alfv\'en velocity near the magnetic sources. This indicates that the use of the computational grid in our simulations is fine enough to trigger plasmoid instabilities shown by the small flux tubes in the CS. Panel (a) displays the pre-existing filament carrying the ring current and no CS is formed at the beginning. At $t=0.52$ in panel (b), the CS fan between the flare and FR is well built up and the first plasmoid marked by the red box appears. The flux tube-like structures of the plasmoid are locally constructed by the twisted field lines. Then at $t=0.96$ in panel (c), the CS is extremely stretched in the middle, and the field lines of the FR crossing the z-axis evolve into a clockwise rotation. At this time, the buffer region under the FR seems more turbulent than the CS, which has only a few blobs near its flanks. Later at $t=1.42$ in panel (d), blobs continue to appear simultaneously close to the flare loop-top and undergo merging or splitting, which makes the internal CS structures quite chaotic. One can find also that the field lines of the FR in the middle rotate at a significant angle with respect to the $yz$-plane, which forms a complex geometry of the CS instead of a flat sheet. Lastly at $t=2.0$ in panel (e), the general CS system becomes completely turbulent at higher altitudes. 
In Figure \ref{tmap}, we plot the spatio-temporal diagram along the z-axis passing the origin using the highest level 5 of AMR, in order to follow the detailed evolution of the CS driven by the ejected FR. In panel (a), one can easily distinguish the fast shock by the shape change and the FR of denser material from the distribution of density in height. The principal X-line (PX), where magnetic reconnection takes place the fastest, is located always close to the lower tip of the CS, causing the energy partition asymmetric. As we can see, the FR undergoes an acceleration phase for $t\lesssim0.52$ and then moves upward with a speed of about $338.5\pm12.8$ km/s. Many blobs seen as fluctuations in density are generated due to instabilities in the CS, and propagate bi-directionally. The reconnection upflows collide with the FR to form a dense buffer region underneath. Unlike the 2D work of \cite{Shen2011,Ye2019}, structures that appear as separate plasmoids might actually be 2D projections of turbulent structures. In panel (b), a pair of termination shocks (TS) are formed above the flare loops and under the FR with the average upstream velocity of about 1028 km/s. Particularly, the 3D interface of TS resembles a thin and long ribbon above the flare loops in our simulation.  From panel (c) and (d), one can see apparent oscillations both in the $x$- and $y$-directions inside the CS and at the location of TSs from the time when blobs appear, unlike the 2D work of \cite{Takahashi2017}. 

\begin{figure}[H]
\centering
\includegraphics[width=9cm,height=7cm,draft=false]{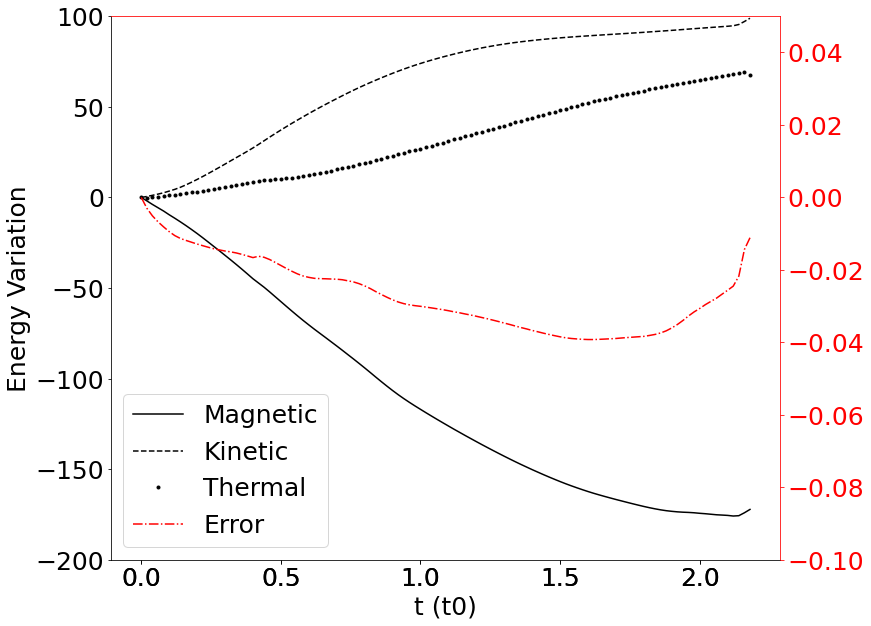}
\caption{\small Time evolution of the variations in total magnetic,  kinetic and thermal energies in the corona (in dimensionless units).  The red dash-dotted line is the error on the total energy estimated by $[\Delta E_M(t)+\Delta E_K(t)+\Delta E_I(t)]/[E_M(0)+E_I(0)]$ \label{energy}}
\end{figure}

By integrating the dimensionless energy equation over the volume $V$ and applying the divergence theorem, we have 
\begin{equation}
\int_V \frac{\partial}{\partial t}(\mathcal{E}+\mathcal{K}+\mathcal{W})dV=\int_V \rho \textbf{v}\cdot\textbf{g}dV-\int_S\left[(\frac{\gamma P}{\gamma-1}+\frac{\rho v^2}{2})\textbf{v}+\textbf{E}\times \textbf{B}+\textbf{F}_C\right]\cdot d\textbf{S},
\end{equation}
where $\mathcal{E}$, $\mathcal{K}$, $\mathcal{W}$ are the thermal, kinetic and magnetic energy densities, respectively. Note that $\textbf{E}=\eta \textbf{J}-\textbf{v}\times \textbf{B}$ is the electric field. The variations of magnetic , kinetic and thermal energies  $\Delta E_M$, $\Delta E_K$, $\Delta E_I$ in the entire simulation domain are given by,
\begin{eqnarray}
&&\Delta E_M = E_M(t)-E_M(0)+\int_0^t\int_S\textbf{E}\times \textbf{B}\cdot d\textbf{S} dt\\
&&\Delta E_K = E_K(t)-E_K(0)-\int_0^t\int_V\rho\textbf{v}\cdot\textbf{g}dVdt+\int_0^t\int_S\left(\frac{\rho v^2}{2}\right)\textbf{v}\cdot d\textbf{S}dt\\
&&\Delta E_I = E_I(t)-E_I(0)+\int_0^t\int_S\left[(\frac{\gamma P}{\gamma -1})\textbf{v}+\textbf{F}_C \right]\cdot d\textbf{S}dt
\end{eqnarray}
Since we are mainly interested in the energy change in the corona and want to reduce the influence from the imperfect handling of the lie-tied condition at the bottom, the above integrals are only calculated for the height $z\geq 0.5$. Figure \ref{energy} shows the time evolution of total magnetic, kinetic and thermal energy variations in the corona, calculated from the data of the AMR level 1. The magnetic energy declines gradually as magnetic reconnection proceeds. And the kinetic energy transferred from the magnetic energy dominates during the entire simulation time, and gradually accumulates and saturates. The generated thermal energy manifests in a piecewise linear phase with the steeper slope after $t\approx 0.52$, corresponding to the end of the acceleration phase. The error on the global energy diagnostic,  estimated by the loss of the initial total energy $[\Delta E_M(t)+\Delta E_K(t)+\Delta E_I(t)]/[E_M(0)+E_I(0)]$, remains smaller than $4\%$ overall. Although 5 AMR levels are utilized in our simulations, the error is computed from the data of the AMR level 1 to save the computational time. Hence, it only gives an upper limit of the numerical error, the actual error could be even smaller, indicating that the simulation runs are quite accurate to respect the energy conservation law. However, some of the total energy is lost due to numerical dissipation.

\begin{figure}[H]
\centering
\begin{minipage}{0.48\linewidth}
\centerline{\includegraphics[width=\textwidth,height=6cm,draft=false]{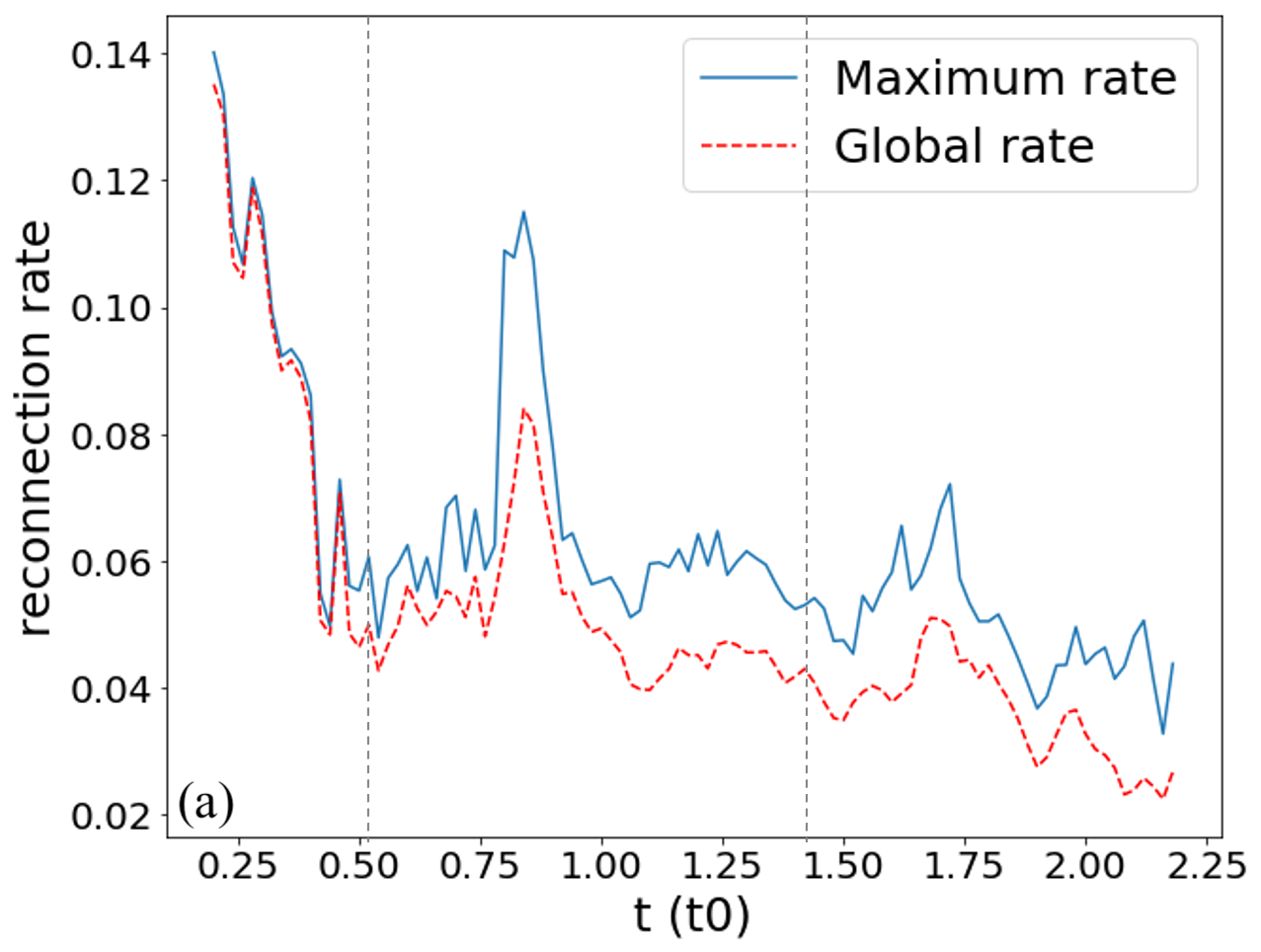}}
\end{minipage}
\begin{minipage}{0.48\linewidth}
\centerline{\includegraphics[width=\textwidth,height=6cm,draft=false]{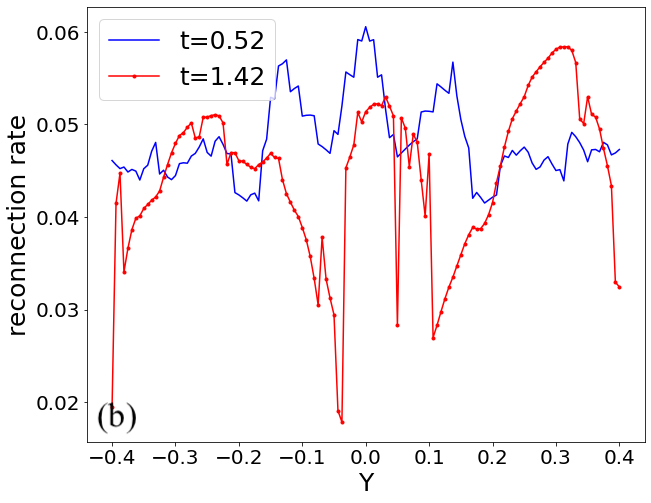}}
\end{minipage}

\caption{\small (a) Magnetic reconnection rates as a function of time;  (b) Local reconnection rates along the $y$-axis at times $t=0.52$ and $1.42$. Vertical dashed lines in panel (a) mark two selected moments for plots in panel (b). \label{rate}}
\end{figure}

Regarding the magnetic reconnection process, we calculated the reconnection rate by estimating the Alfv\'enic Mach number in the inflow region near the PX-line in the 3D configuration \citep{Forbes2000,Jiang2021}. To do so, we compute the global rate from the average inflow velocity and the Alfv\'en speed, and the maximum rate along the PX-line. In Figure \ref{rate}(a), we present the global and maximum rates for time $t\geq 0.2$, when the sheet-like CS is well formed between the CME and the flare. As we can see, the global rate decreases in the impulsive phase (uniform stage) for $t<0.5$ and then oscillates and decays at the same time (nonuniform stage), which is consistent with the observation work of \cite{Song2018}. In the first stage, the reconnection process is externally driven by the FR motion, resulting in a large reconnection rate greater than 0.1 at the very beginning. Later in the second stage, the self-organized plasmoid or turbulent reconnection might play a competing role causing peaks in the reconnection rates. The global and maximum rates are close for $t<0.5$, indicating that it is in the linear phase with almost the same magnitude, while they deviate from each other for $t \geq0.5$ because of the appearance of plasmoids. In Figure \ref{rate}(b), we plot also the local reconnection rate at each PX-point along the $y$-axis at times $t=0.52$ and $1.42$. At $t=0.52$, the reconnection rate greatly increases between y=-0.18 and y=0.18, where the first plasmoid is just generated (See Figure \ref{yzplane}). At $t=1.42$, the CS becomes very turbulent with the appearance of many plasmoids, and the rate is enhanced along the PX-line by different amplitudes depending on the location of these plasmoids (See also Figure \ref{yzplane}). In other words, every PX-point along the PX-line can have a different reconnection rate, and turbulence normally brings in a higher rate in local area for a given time. 

\subsection{Fine structures and processes of the CS}
\begin{figure}[H]
\centering
\begin{minipage}{0.3\linewidth}
\centerline{\includegraphics[width=\textwidth,height=6cm,draft=false]{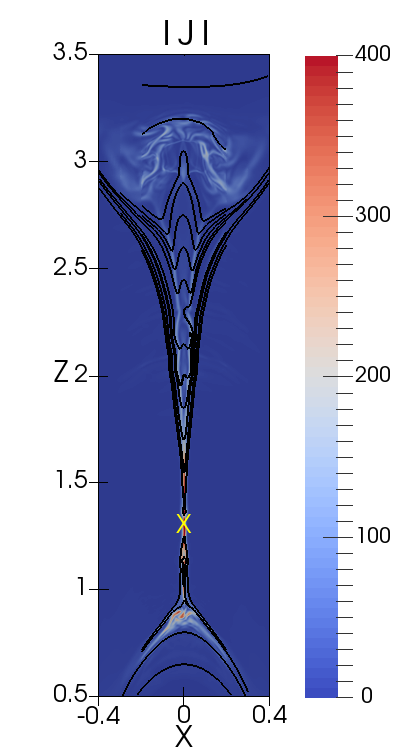}}
\end{minipage}
\begin{minipage}{0.3\linewidth}
\centerline{\includegraphics[width=\textwidth,height=6cm,draft=false]{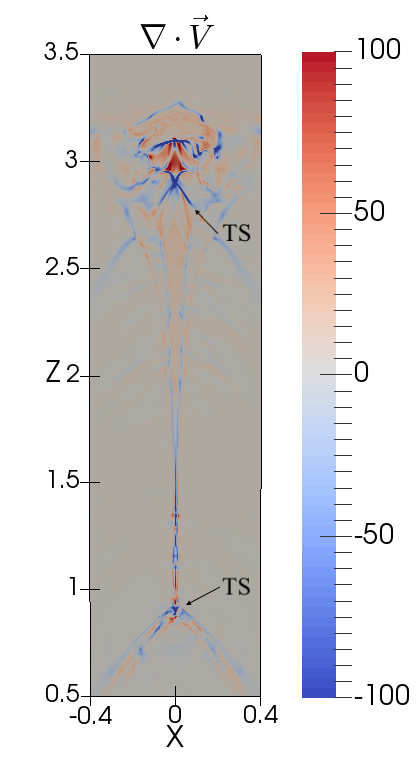}}
\end{minipage}
\begin{minipage}{0.3\linewidth}
\centerline{\includegraphics[width=\textwidth,height=6cm,draft=false]{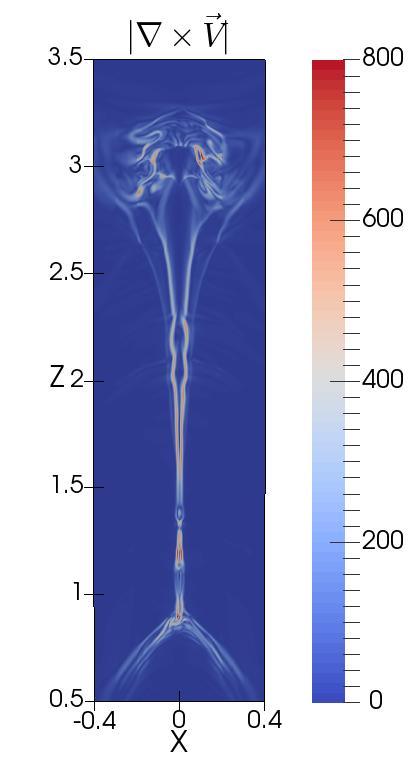}}
\end{minipage}
\centerline{(a) Vertical slices at the center $y=0$}
\vfill
\begin{minipage}{0.18\linewidth}
\centerline{\includegraphics[width=\textwidth,height=5cm,draft=false]{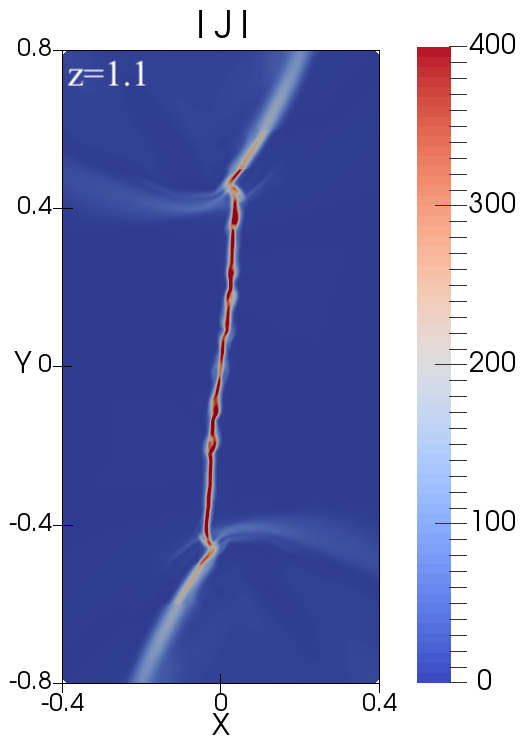}}
\end{minipage}
\begin{minipage}{0.18\linewidth}
\centerline{\includegraphics[width=\textwidth,height=5cm,draft=false]{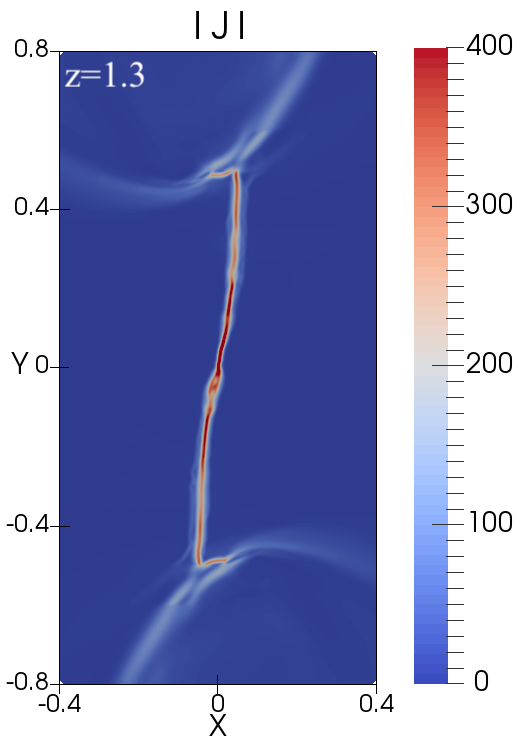}}
\end{minipage}
\begin{minipage}{0.18\linewidth}
\centerline{\includegraphics[width=\textwidth,height=5cm,draft=false]{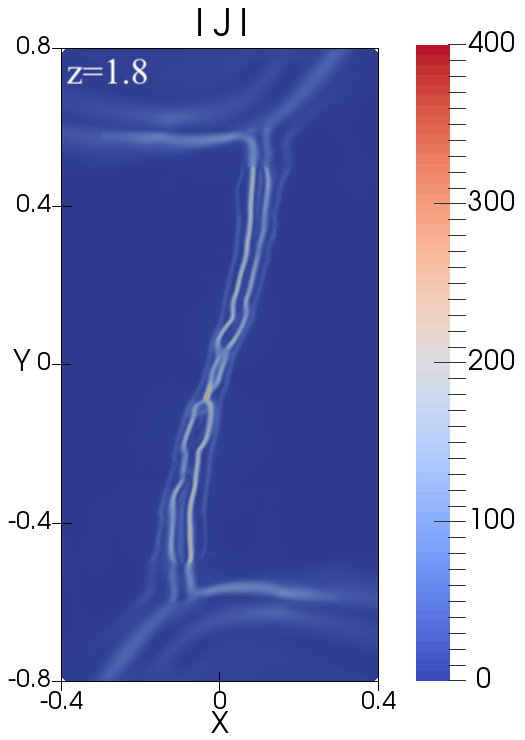}}
\end{minipage}
\begin{minipage}{0.18\linewidth}
\centerline{\includegraphics[width=\textwidth,height=5cm,draft=false]{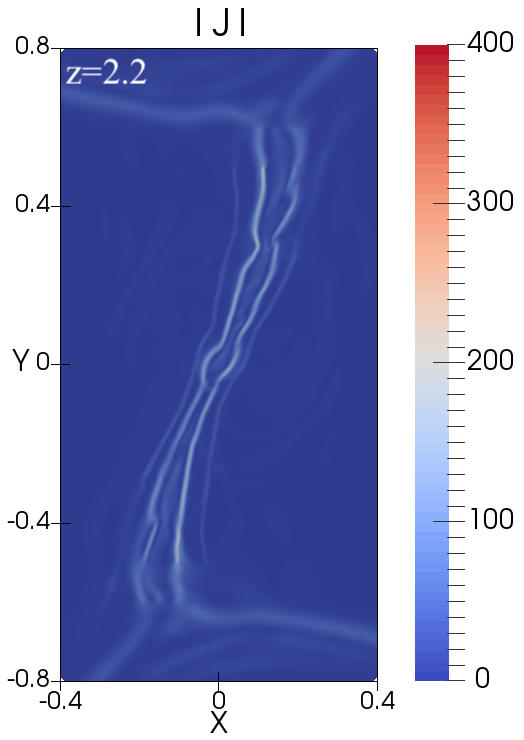}}
\end{minipage}
\begin{minipage}{0.18\linewidth}
\centerline{\includegraphics[width=\textwidth,height=5cm,draft=false]{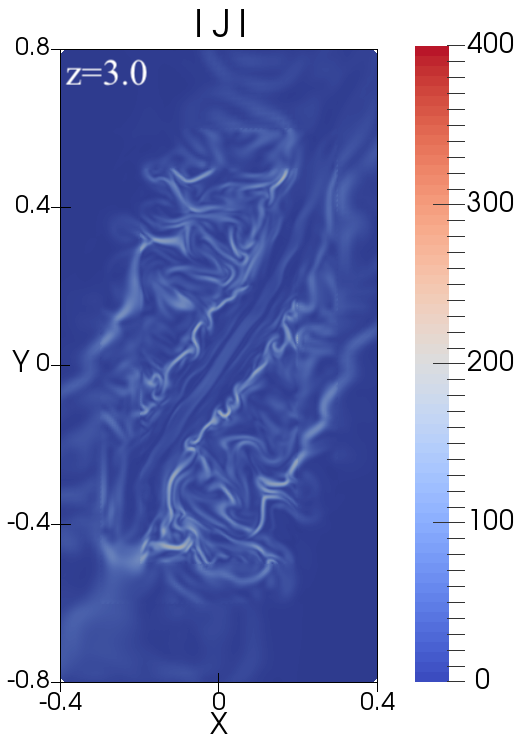}}
\end{minipage}
\centerline{(b) Horizontal slices at different heights}
\caption{\small (a) Distributions of the current density $|J|$, divergence $\nabla\cdot \vec{V}$ and velocity curl $|\nabla\times \vec{V}|$ on the xz-plane at $t=1.42$. (b) Current density on the xy-plane at different heights. The black solid lines in the current density distribution in panel (a) are the magnetic field and the yellow cross `X' therein shows the PX-point on this slice. \label{cut} }
\end{figure}

In this section, we will focus on the small-scale fine structures in the 3D CS formed during the solar eruption. For instance, one can see that the global CS layers have already turned into the very turbulent state at $t=1.42$ in Figure \ref{global}. At this time, we display the planar cuts of different parameters in Figure \ref{cut}. Panel (a) presents the distributions of the current density, divergence and velocity curl on the vertical slice at the center ($y=0$). The CS below the PX-point stays extremely thin due to the high pressure nearby, and the magnetic field displays a Sweet-Parker configuration. Meanwhile, the magnetic field above the PX-point forms the 'W' shape and the plasma flow presents a classical Petschek configuration with a pair of the slow-mode shocks (SSs). The existence of plasmoids thickens the local CS width and makes the reconnection process even more complex. By following the work of \cite{Wang2009}, the divergence and the curl of velocity are useful information for identifying the location of fast-mode shocks (FSs) and SSs, respectively. One can notice particularly that a pair of TSs form the valleys of the plasma at the tips of the CS, where the divergence takes the minimum value. The strong compression of the plasma occurs at the location of the TSs, and the interface of the upper TS is much larger than the lower one. Also , SSs are generated from the PX-point and propagate bi-directionally along the field lines, which can be identified from Rankine-Hugoniot relations \citep{Shiota2005,Mei2020}. Notice that the flare loop-top and the buffer region under the FR are full of the SSs, which are the important contributors to the local plasma heating. Figure \ref{cut}(b) displays the horizontal slices of the current density at different heights of the CS. As we can see, the two ends of the CS in the $xy$-plane are connected to the arms of the FR. At the low altitudes of the CS at $z=1.1$ and $z=1.3$, the thickness of the CS is very thin, but it can be broadened locally by the existence of the blobs. At the high altitudes at $z=1.8$ and $z=2.2$, the thickness of the CS is larger and becomes very nonuniform in the $y$-direction. Lastly in the buffer region at $z=3$, the CS is extremely disrupted by the turbulence, which is triggered by the Rayleigh-Taylor instability suggested by \cite{Xie2022}. From lower to higher position, the fan of the CS becomes more and more tilted and forms an inverse 'S' structure for the upper CS due to the global deflection of FR during the eruption. \par
\begin{figure}[H]
\centering
\includegraphics[width=12cm,height=18cm,draft=false]{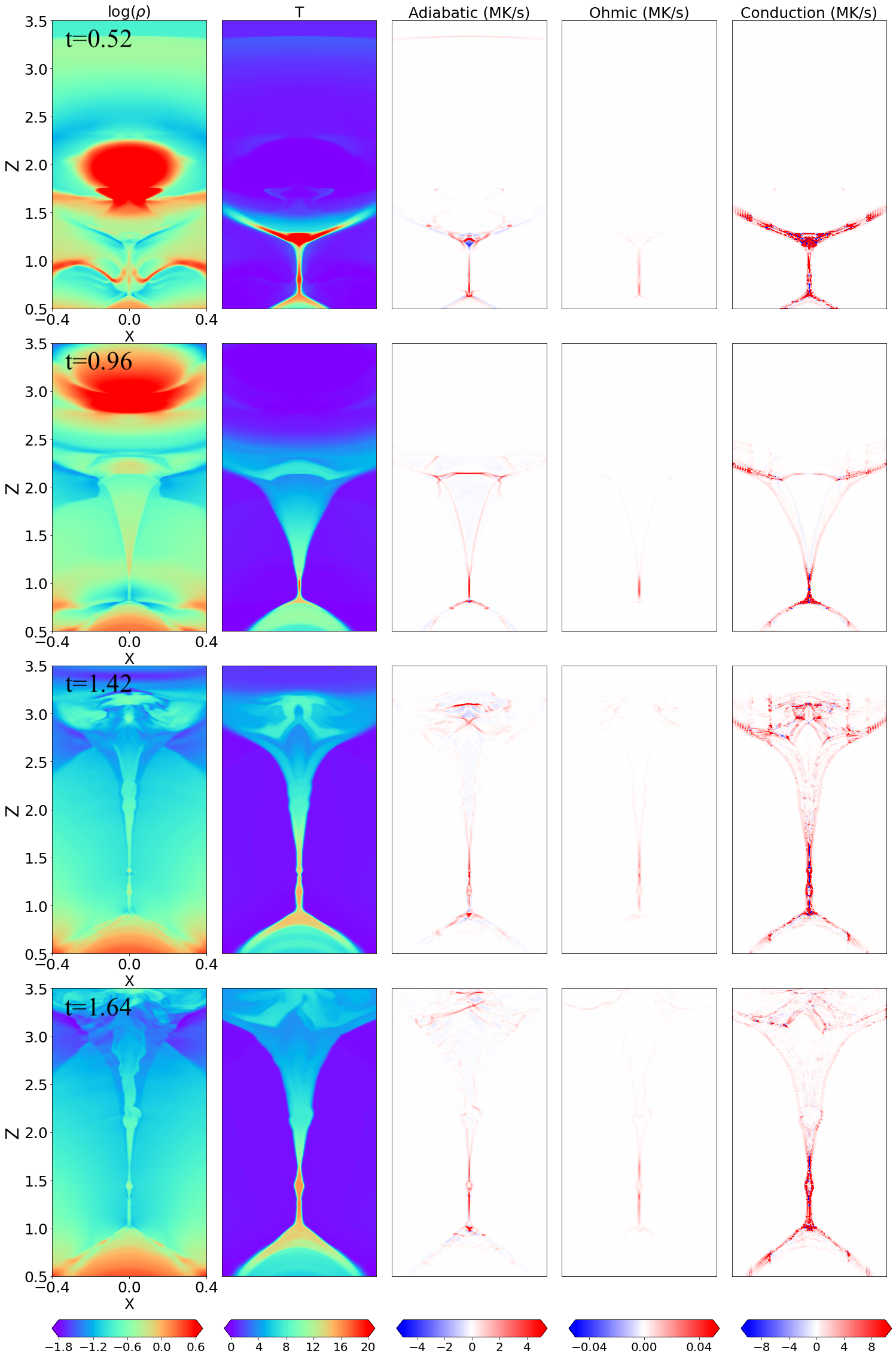}
\caption{\small Log of density (first column), temperature (second column), adiabatic term (third column), ohmic heating (fourth column) and thermal conduction (fifth column) for the $xz$-plane at the center ($y=0$) at different times. Note that the contribution from adiabatic, ohmic heating and thermal conduction is shown in cgs units for better understanding the plasma heating process in the current sheet.  \label{heat2D}}
\end{figure}

\begin{figure}[H]
\centering
\includegraphics[width=16cm,height=10cm,draft=false]{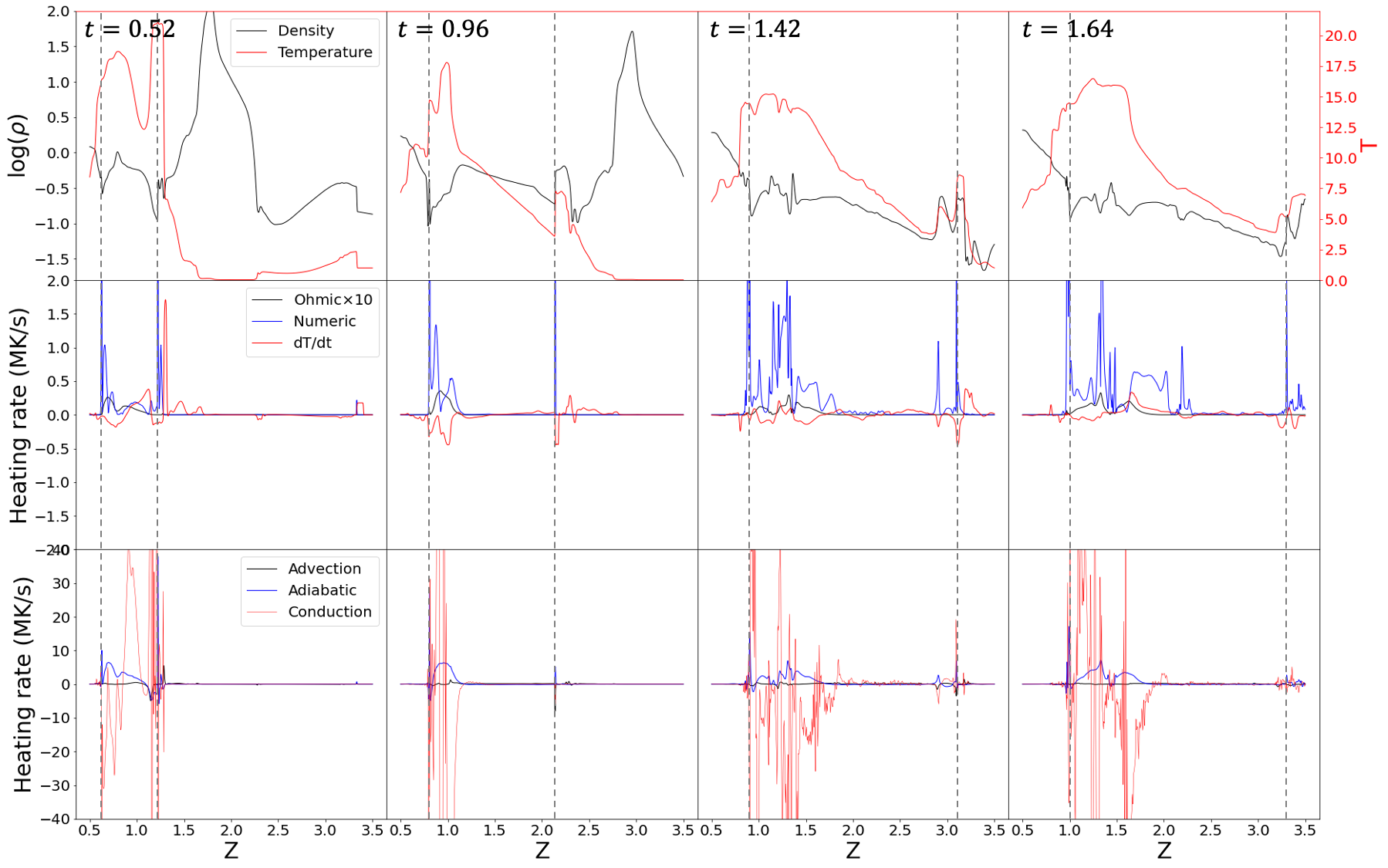}
\caption{\small Plots of quantities along the central line (x=0) in Figure \ref{heat2D} at the same times as in this figure: density and temperature (first row), heating terms and change in the temperature ($dT/dt$) and heat transport terms. The vertical dashed lines mark the current sheet ends. \label{heat1D}}
\end{figure}

   
In order to understand how the plasma is heated in the CS, we rewrite the energy equation as:
\begin{equation}\label{heateq}
\frac{\partial T}{\partial t}=-\textbf{v}\cdot\nabla T-(\gamma-1)T\nabla\cdot \textbf{v}+\frac{\gamma-1}{\rho}(-\nabla\cdot\textbf{F}_C+\eta J^2),
\end{equation}
where the terms $-\textbf{v}\cdot \nabla T$ and $-(\gamma-1)T\nabla\cdot \textbf{v}$ are the advection and adiabatic terms, $-\nabla\cdot\textbf{F}_C$ is the thermal conduction term, and $\eta J^2$ is the ohmic heating term. The density, temperature and different heating terms on the $xz$-plane at the center ($y=0$) are shown in Figure \ref{heat2D}. Here, the calculations are based on the data of the AMR level 5, and the sampling period is $0.02t_0$. At the early time $t=0.52$, the density and temperature distributions show a clear Y-structure above the flare loops, and the hot dense CS becomes extremely thin. The adiabatic term is primarily positive inside the CS due to compression of the plasma by the strong reconnection inflows, except for its upper end. The ohmic heating is at least two orders smaller than the adiabatic one, and conduction is a dominating contributor against the other heating terms in the CS, at the flare loop-top and around the CME bubble. At $t=0.96$, more plasma accumulates in the upflow region to make the upper CS width larger, where the density and temperature distributions become quite smooth. The adiabatic term is only large in the lower CS and at two ends of the global CS due to the compression by TSs. The ohmic heating term is most important in the lower CS. Thermal conduction spreads the temperature out of the CS and produces a “thermal halo” around it \citep{Seaton2009}. This can be seen in the hot temperature structure, which is much wider than the dense region at the same location. Later at $t=1.42$ and $1.64$, the CS becomes long and thin, stretched by the rising FR, and many hot and dense plasmoids or blobs can be seen therein. At these times, the adiabatic and thermal conduction terms fragment into small pieces. From the density distribution, one can see the tuning fork structures formed above the flare loops and at bottom of the CME with two arms highly compressed, where the adiabatic and conduction effects are still significant. The ohmic term remains ignorable compared with other heating terms. We note that the conduction term is strongly concentrated in regions with high temperature gradients, indicating that the AMR resolves these regions. For instance, the thin heating regions at $t=0.96$ at $z=1.5$ are resolved into 52 grid cells by AMR. As discussed in Figure \ref{energy}, the total energy is lost due to numerical dissipation in terms of the viscous heating. In order to quantify the loss of energy, the numerical kinematic viscosity coefficient due to the upwinding scheme as \cite{Reeves2019} is estimated with the equation
\begin{equation}\label{viscoeff}
\nu_N=\sum_{i,j,k}\frac{v_i\Delta x_i}{2}(1-v_i\frac{\Delta t}{\Delta x_i}),
\end{equation}
where $\Delta x_i$ is the grid spacing for the $i$th dimension, i.e. ($\Delta x, \Delta y, \Delta z$), $v_i$ is the plasma velocity in the cell in the $i$th dimension, and $\Delta t$ is the time step in the code. Then the excess viscous heating produced by the numerical diffusion reads as 
\begin{equation}
H_{\textrm{nvisc}}=\rho\nu_N\left[\frac{1}{2}e_{ij}e_{ij}-\frac{2}{3}(\nabla\cdot\textbf{v})^2\right],
\end{equation}
where $e_{ij}$ is the rate-of-strain tensor. On the other hand, we roughly measure the kinematic viscous coefficient $\nu_m$ by adding the numeric term $(\gamma-1)H_{\textrm{nvisc}}/\rho$ into the right side of Eq.(\ref{heateq}) as follows:
\begin{equation}
\nu_m = \frac{|\frac{dT}{dt}+\textbf{v}\cdot\nabla T+(\gamma-1)T\nabla\cdot\textbf{v}-\frac{\gamma-1}{\rho}(-\nabla\cdot\textbf{F}_C+\eta J^2)|}{(\gamma-1)\left[\frac{1}{2}e_{ij}e_{ij}-\frac{2}{3}(\nabla\cdot\textbf{v})^2\right]}.
\end{equation}
For comparison, $\nu_N$ and $\nu_m$ are averaged over the region with $(x,z)\in[-0.2,0.2]\times[0.5,3.5]$ for $t=1.42$ and $1.64$ in Figure \ref{heat2D} to suppress the unnecessary errors, which both give the value close to 0.0035. This confirms that Eq.(\ref{viscoeff}) is a reliable way to quantify the numerical diffusion. Figure \ref{heat1D} plots the density, temperature, together with the heating terms along the central line ($x=0$) in Figure \ref{heat2D}. The ohmic term is multiplied by 10 to be visible. Here, we also plot the numerical viscous heating. At $t=0.52$, the density peaks at the core of the CME and the CS length is short. The temperature in the CS reaches over 20 MK and is highest at the upper end. The $dT/dt$ is mostly negative in the CS, but it becomes positive and relatively strong near the upper end of the CS, indicating strong heating at the CME bottom. The numerical diffusion is shown to be generally greater than the ohmic term with the two peaks at the ends of the CS. The adiabatic term at this time is the main contributor for heating in the CS. Conduction is the dominating factor that cools the lower half CS and transfers the heat to the bottom of the CME. Advection plays a similar role to conduction but with the magnitude much smaller. At $t=0.96$, the CS is stretched longer by the rising FR and the super hot structure ($>10$ MK) concentrates only in the lower CS between $z=0.8$ and $z=1.1$. All heating terms including the numerical diffusion behave similarly as the previous time except that they are more smooth in the upper CS for $z>1.1$. At $t=1.42$ and $1.64$, the CS enters the fragmented and fully turbulent phase due to plasmoid instability \citep{Shen2011,Ye2019}. The temperature of the CS reaches 17.5 MK and the range of temperatures is relatively narrow as observed in \cite{Warren2018}. The hot and dense plasmoids or blobs start to disrupt the distributions of the density and temperature in the CS. The numerical diffusion has become significant due to the large local velocity or velocity changes produced by the instabilities in the CS, indicating a higher viscous heating rate around the plasmoids or blobs. The adiabatic term remains mostly positive in the lower half of CS due to the strong compression by the reconnection inflows. And the advection term mainly plays a role for cooling the plasma against the adiabatic term. Thermal conduction serves to smooth out the temperature distribution along the CS, and the large magnitude of this term is mainly located in the lower half CS. It implies that substantial heat is conducted to the lower end of the CS to make the flare loops hotter rather than the CME.

\begin{figure}[H]
\centering
\begin{minipage}{0.9\linewidth}
\centerline{\includegraphics[width=\textwidth,height=5cm,draft=false]{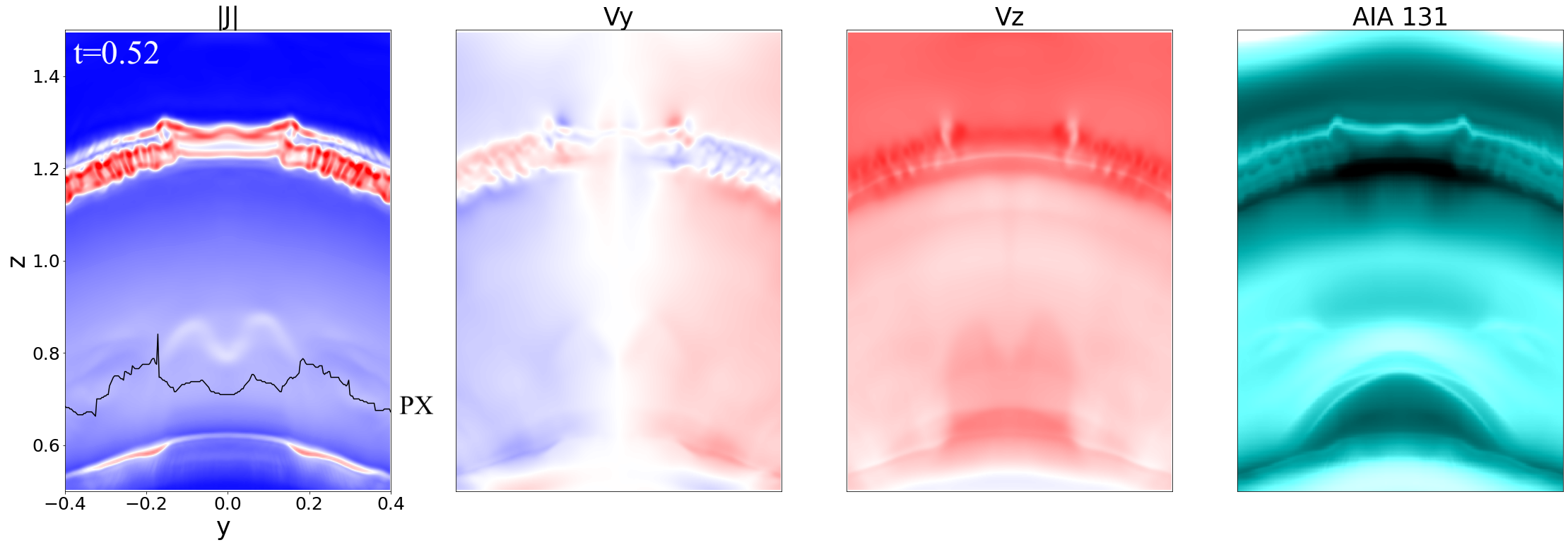}}
\end{minipage}
\vfill
\begin{minipage}{0.9\linewidth}
\centerline{\includegraphics[width=\textwidth,height=5cm,draft=false]{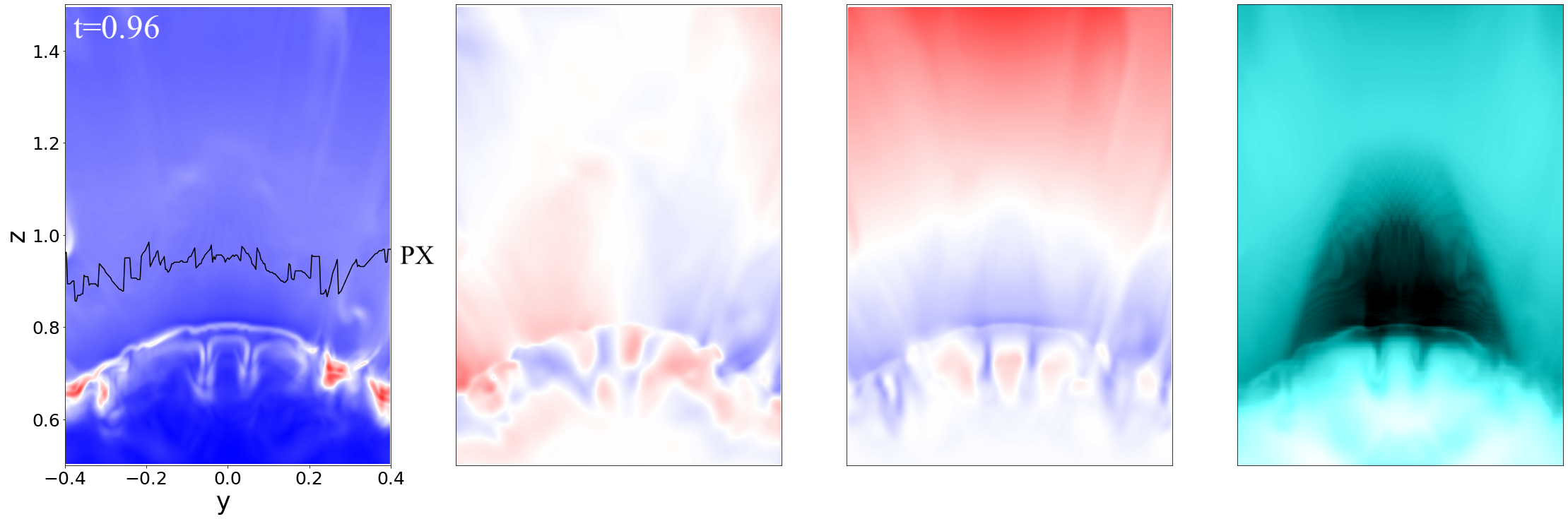}}
\end{minipage}
\vfill
\begin{minipage}{0.9\linewidth}
\centerline{\includegraphics[width=\textwidth,height=6cm,draft=false]{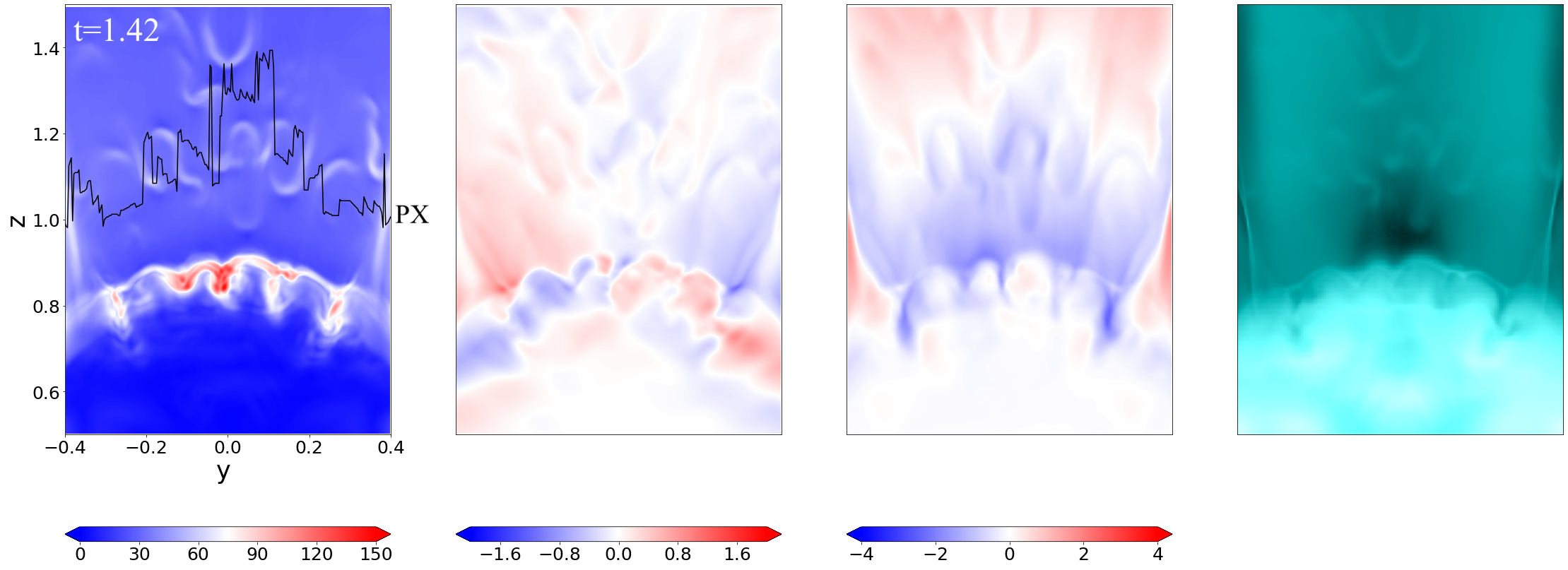}}
\end{minipage}
\caption{\small Face-on view of the CME-flare current sheet at given times $t=0.52, 0.96, 1.42$. From left to right for each row, there are the current density $|J|$, velocities $V_y, V_z$ in the y- and z-directions, and synthetic EUV image for AIA 131 $\mathring{A}$. The black solid lines in the first column are the PX-lines. Note that $|J|,V_y,V_z$ are averaged over the LOS for $x\in [-0.1,0.1]$. \label{yzplane} }
\end{figure}

We present the face-on view of the region of interest including the flare and a part of CS at the center plane (x=0) for selected times in Figure \ref{yzplane}. The physical quantities shown are averaged along the line of sight (LOS) for the length of $x\in [-0.1,0.1]$. For the early stage of the eruption at $t=0.52$, the first plasmoid just appears and the reconnection outflows are closely uniform. The y- and z-direction velocity fields oscillate slightly in the region under the FR rather than the loop-top region. The flare structure in the synthetic AIA 131$\mathring{A}$ map is smooth, while the CS shows a cavity inside due to the low density region formed by the reconnection inflows. Later at $t=0.96$, the current distribution at the flare loop-top becomes turbulent, where oppositely directed plasma outflows appear. Unlike the 2D cases \citep{Ye2019,Ye2020}, the reconnection outflows in 3D are not only along the flare arcades in the xz-plane, but also have the apparent movement in y-direction. The mixture of negative and positive velocities in the $y$- and $z$-directions at the interface region of the strong electric current indicates that turbulent reconnection may readily occur. Accordingly in the channel of AIA 131$\mathring{A}$, the finger-like structures, known as SADs, can be observed between the post-flare loops and the loop-top, which are formed due to the Rayleigh-Taylor or Richtmyer-Meshkov instabilities \citep{Shen2022}. Then at $t=1.42$, the reconnection engages into the nonlinear phase with highly nonuniform velocity field in the CS. One can easily recognize the blobs in different flux tube substructures from the current distribution in the CS, and the velocity distribution of each blob can be also different. If one blob particularly has both negative and positive values of $V_z$, the splitting process occurs, which cannot be seen in 2D cases \citep{Ye2019,Zhang2022}. In the AIA 131$\mathring{A}$ image, the height of SADs increases with the accumulated flare arcade, and the moving blobs are slightly brighter than the background CS fan, which might extend the spikes of SADs to a higher place above the loop-top. From the first column of this figure, one can see that the PX position is never at the same height along the $y$-axis, especially that the complex structures of blobs can strongly disrupt the primary reconnection sites. \par
\begin{figure}[H]
\centering
\includegraphics[width=0.8\textwidth,height=12cm,draft=false]{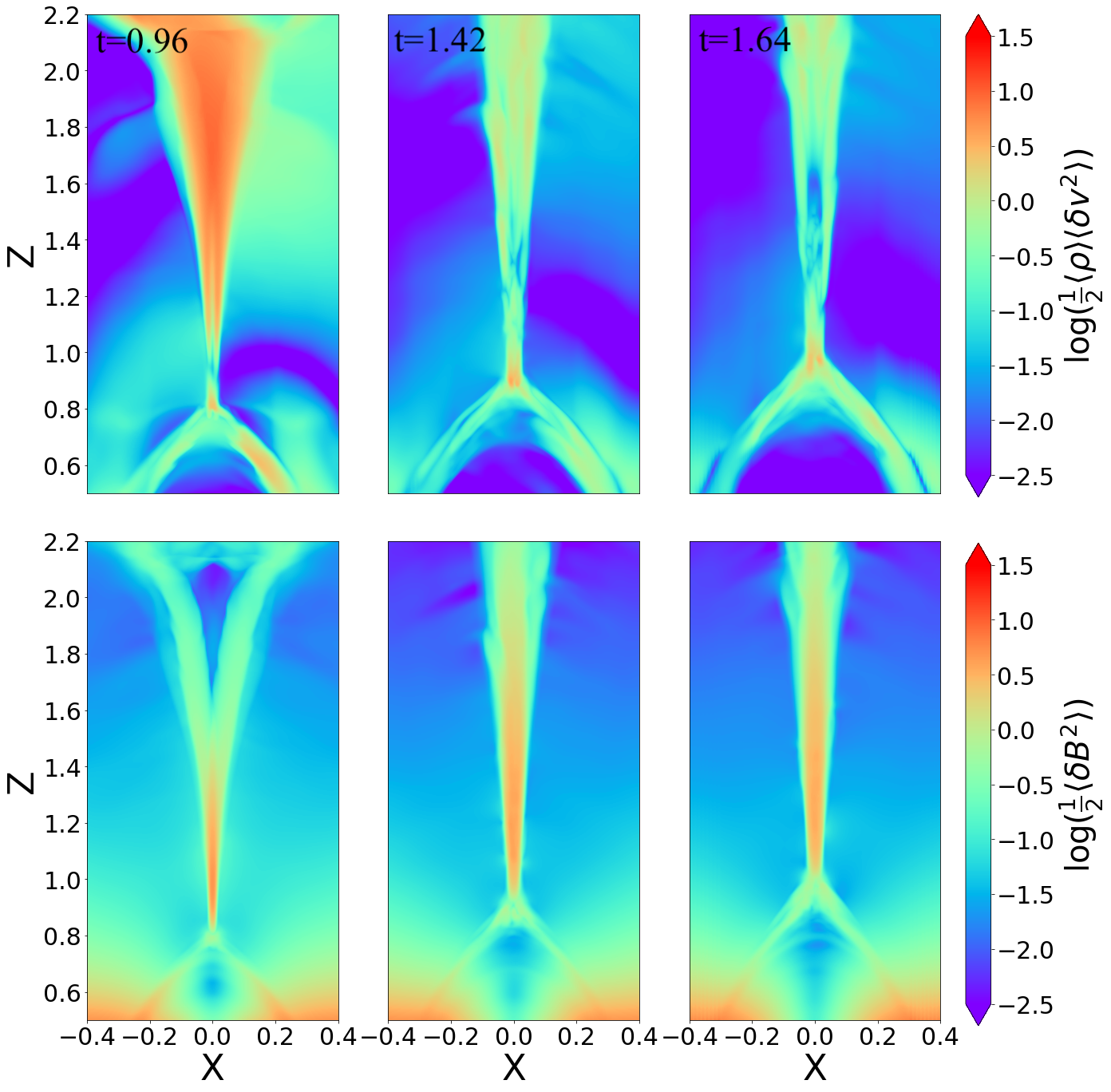}
\caption{\small Distributions of the turbulent kinetic and magnetic energy density at different evolution times $t=0.96$, 1.42 and $1.64$.  \label{turb} }
\end{figure}

Furthermore, we study the turbulence amplitudes in space by estimating the turbulent kinetic and magnetic energy densities at different times. The velocity and magnetic fluctuations $\delta V$ and $\delta B$ are defined as, respectively,
\begin{eqnarray}
\delta V(t,x,y,z)^2 = (V_x(t,x,y,z)-\langle V_x\rangle(t,x,z))^2\nonumber\\
+(V_y(t,x,y,z)-\langle V_y\rangle(t,x,z))^2\nonumber\\
+(V_z(t,x,y,z)-\langle V_z\rangle(t,x,z))^2,
\end{eqnarray}
and 
\begin{equation}
\delta B^2=(B_x-\langle B_x\rangle)^2+(B_y-\langle B_y\rangle)^2+(B_z-\langle B_z\rangle)^2,
\end{equation}
where $\langle \cdot \rangle$ is taken as the average value in the $y$-direction for the region of interest. Particularly, we choose here to compute the average value for a physical quantity over the length for $y\in[-0.2,0.2]$. Then the turbulent kinetic energy density is given by $<\rho><\delta V>^2/2$, and the turbulent magnetic energy density is $<\delta B>^2/2$. Figure \ref{turb} shows the spatial distributions of the corresponding turbulent energy densities at different times. At $t=0.96$, the turbulent component of kinetic energy generally dominates over that of magnetic energy in the CS except the region very close to the PX-line (see Figure \ref{yzplane}). Then the turbulent kinetic energy density becomes smaller and breaks up into small pieces at $t=1.42$ and 1.64. At these times, the turbulent components of kinetic and magnetic energies are comparable, and their distributions with height are highly inhomogeneous. We notice that both energy densities somewhat enhanced at the loop-top region, resulting from the interchange instabilities of backflows therein \citep{Takasao2016}. And the turbulence of kinetic energy is shown to be the strongest at the loop-top, which is consistent with the observational work of \cite{Kontar2017,Stores2021}.

\begin{figure}[H]
\centering
\begin{minipage}{0.23\linewidth}
\centerline{\includegraphics[width=\textwidth,height=4.5cm,draft=false]{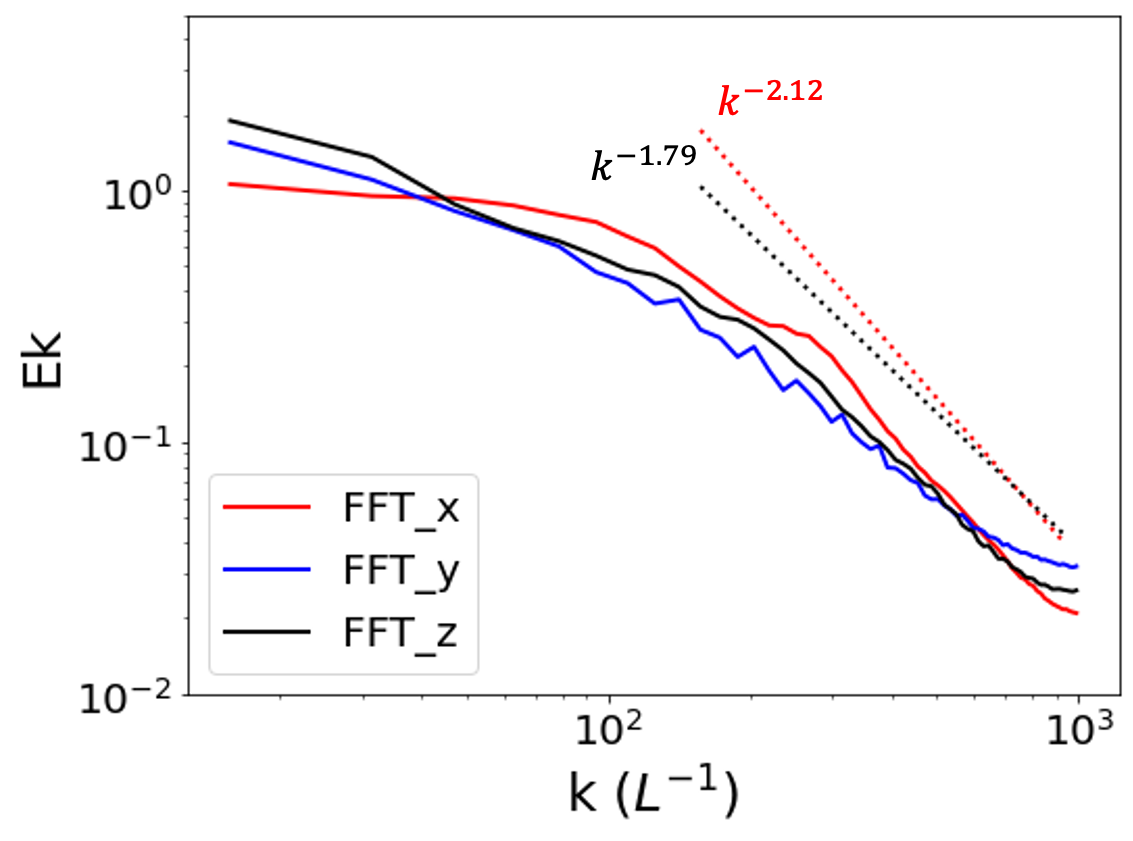}}
\centerline{A: In the flare loops}
\end{minipage}
\begin{minipage}{0.23\linewidth}
\centerline{\includegraphics[width=\textwidth,height=4.5cm,draft=false]{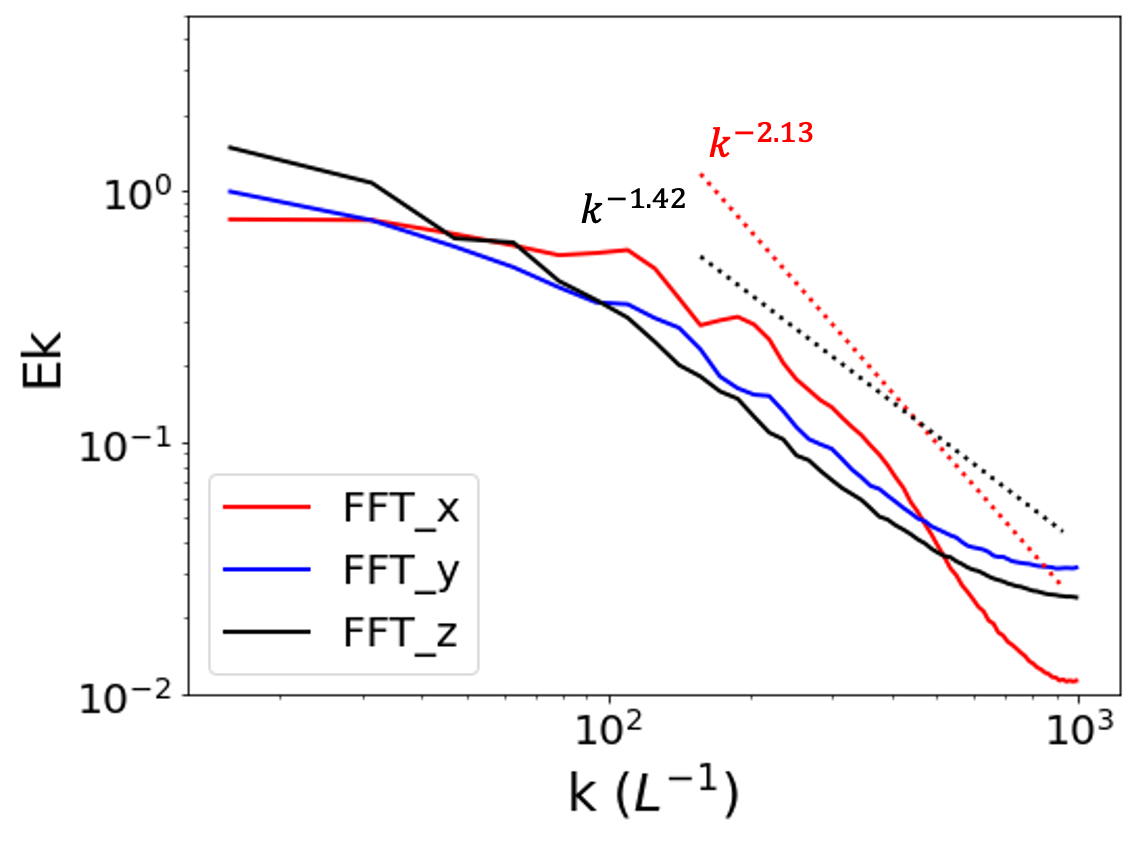}}
\centerline{B: close to the PX}
\end{minipage}
\begin{minipage}{0.23\linewidth}
\centerline{\includegraphics[width=\textwidth,height=4.5cm,draft=false]{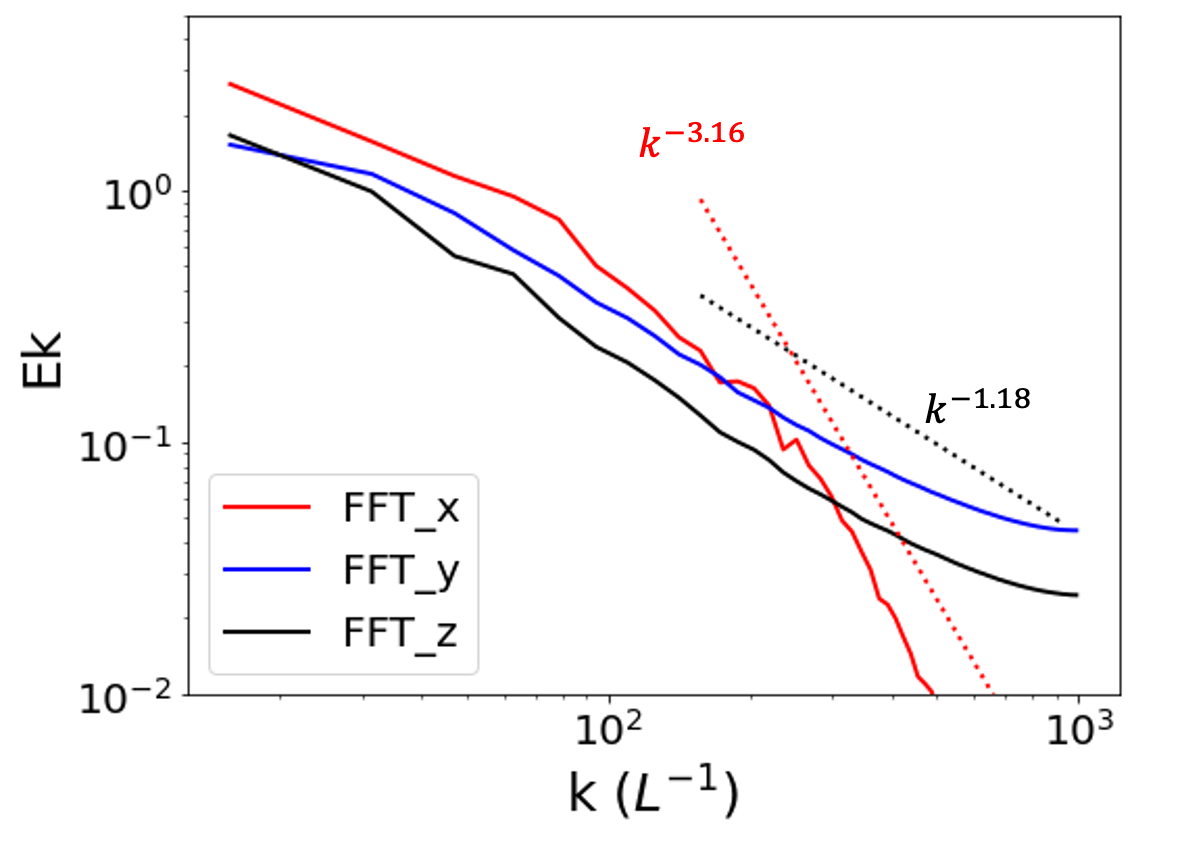}}
\centerline{C: far above the PX}
\end{minipage}
\begin{minipage}{0.23\linewidth}
\centerline{\includegraphics[width=\textwidth,height=4.5cm,draft=false]{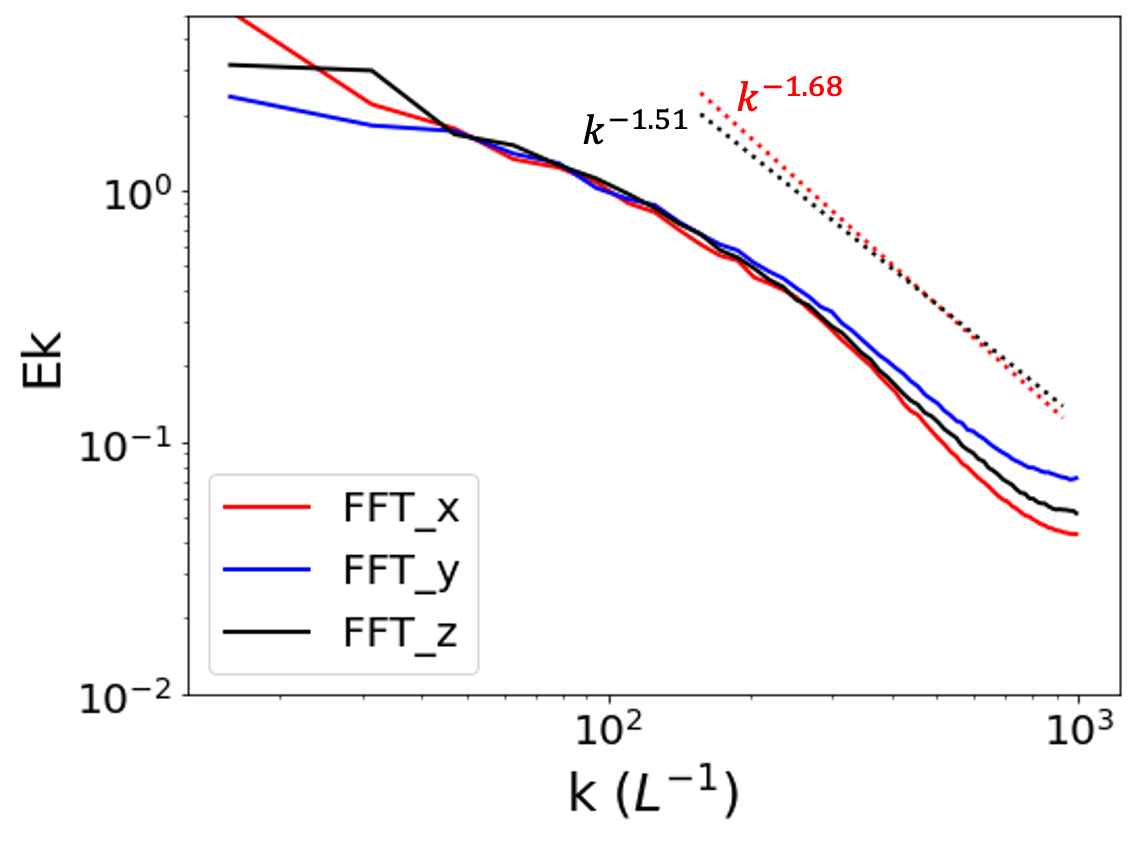}}
\centerline{D: under the CME}
\end{minipage}
\caption{\small Fourier power spectra of the velocity in the $x-$, $y-$ and $z-$directions for different parts of the CS as a function of the wavenumber $k$. The red and black dotted lines in each panel denotes the spectral slopes in the $x$- and $z$-directions for different regions. \label{spec3D} }
\end{figure}

Additionally, it is worthwhile to quantify the anisotropy of turbulence at different heights of the CS. By picking up the simulation data at $t=1.42$ (see Figure \ref{global}), we select four cubic boxes A-D in four specific regions: in the flare loops, close to the PX, far above the PX and under the CME. The boxes are defined as A: $[-0.2,0.2]\times[-0.2,0.2]\times[0.6,1]$, B: $[-0.2,0.2]\times[-0.2,0.2]\times[1,1.4]$, C: $[-0.2,0.2]\times[-0.2,0.2]\times[1.8,2.2]$ and D: $[-0.2,0.2]\times[-0.2,0.2]\times[2.8,3.2]$ in $(x,y,z)$ coordinates. Figure \ref{spec3D} shows the average Fourier power spectra obtained from the velocity field for each box as a function of the wavenumber $k$, normalized by the characteristic length $L$. The power spectra are calculated using 3D Fourier transform \citep{Kirby2005} and then the spectrum in each direction is obtained by summing up the power coefficients in the other directions. The associated Fourier transforms show kinetic structures in the $x$-, $y$- and $z$-directions, respectively. In box A, B and C, the power in the $y$- and $z$-directions show reasonable power law behavior from 100 or 200$L^{-1}$ to 700 or 800$L^{-1}$. The slopes are flatter than the dotted lines associated with the $x$-direction. The power in the $x$-direction seems to be curved rather than power law, or to consist of several narrow power law intervals. Box B pertains to the X-line region where turbulence is strongly driven. It is complicated by the fact that the CS is narrower than the box, so $x$-direction power at small $k$ is suppressed. So the $x$-direction might not have a clear inertial range, but there is an excess of power on scales of the roughly CS thickness, which varies with the height $z$. Box D clearly shows well-developed, isotropic turbulence. In this case the slopes in the power law portions at small scales seem to agree with the range -1.5 to -1.7 expected for Goldreich-Sridhar or Kolmogorov turbulence \citep{Beresnyak2017}. That turbulence is probably driven by the interaction between the reconnection outflow and the erupted flux rope, as described in \cite{Ye2021}. That could account for the higher overall level of power at all scales. There has been considerable work on the power law indices and the inner and outer scales of turbulence in simulations of current
sheets \citep{Barta2011, Shen2013, Boldyrev2006, Boldyrev2020, Kowal2017,Edmondson2017, Zharkova2021, Ye2019, Ye2020}. Given the limited inertial range and the gradual character of the change in slope at large wavelengths in our simulations, there is some uncertainty in both the spectral index and the outer scale.  \par
If the $x$-direction turbulence in box C has evolved from that in box B, the higher power at large scales could be partly due to the larger CS thickness far from the X-point. The smaller power at small scales could result from mergers of islands. It seems too steep to result from cascade to the dissipation scale. In box A in the flare loops, one could also imagine that the stronger power at large scales comes about because the CS is thicker. There may be an inertial range with a somewhat steeper slope than the $y$- and $z$-directions. The TS will strongly affect the turbulence, as in \cite{Shen2018}, but the spectra are similar enough to the spectra near the X-line that they may be strongly influenced by the turbulence advected from above.

\begin{figure}[H]
\centering
\subfigure[$t=1.38$]{\includegraphics[width=0.22\textwidth,height=6cm,draft=false]{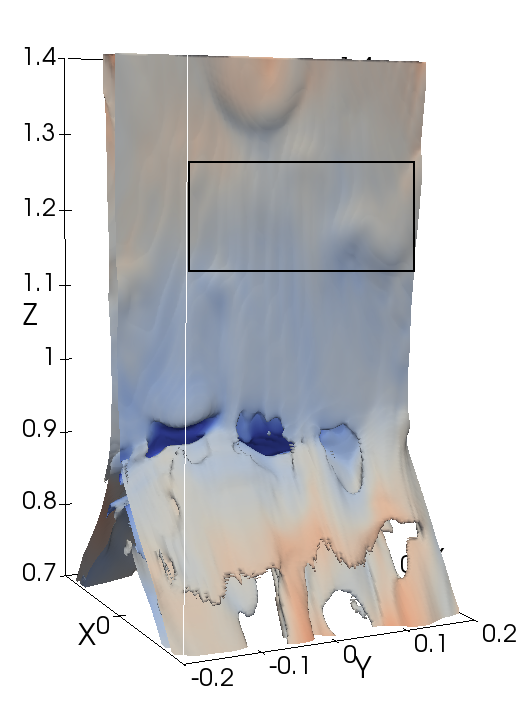}}
\subfigure[$t=1.40$]{\includegraphics[width=0.22\textwidth,height=6cm,draft=false]{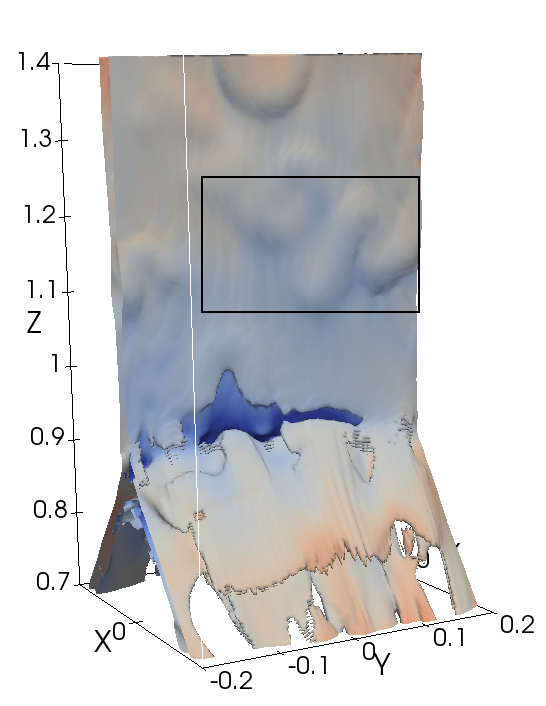}}
\subfigure[$t=1.42$]{\includegraphics[width=0.22\textwidth,height=6cm,draft=false]{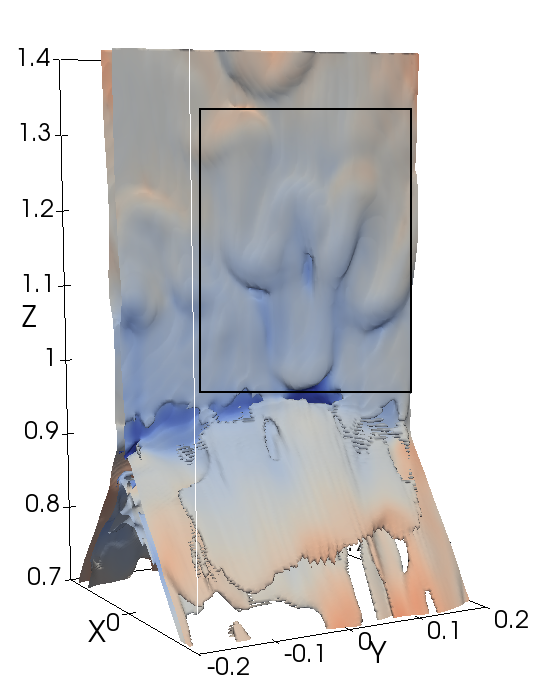}}
\subfigure[$t=1.44$]{\includegraphics[width=0.28\textwidth,height=6cm,draft=false]{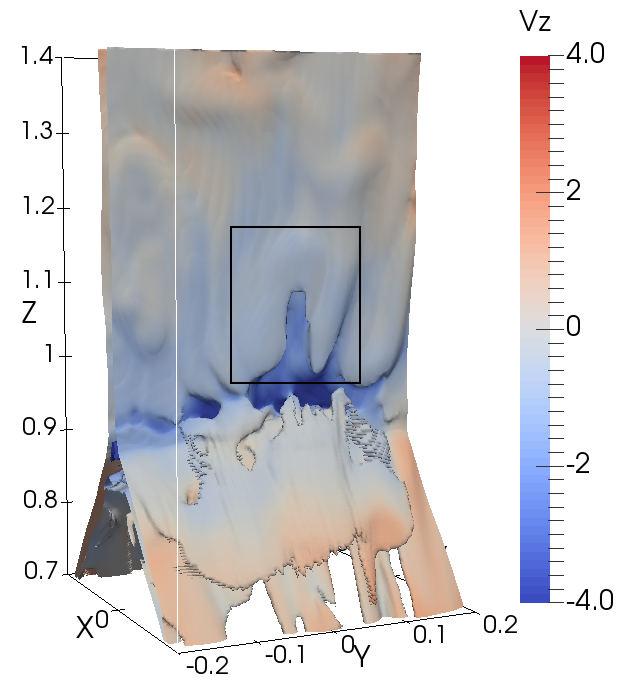}}
\subfigure[$t=1.38$]{\includegraphics[width=0.22\textwidth,height=6cm,draft=false]{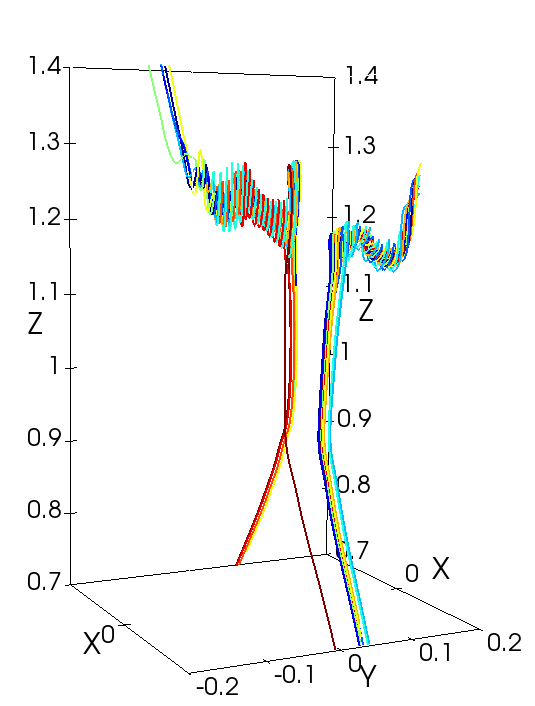}}
\subfigure[$t=1.40$]{\includegraphics[width=0.22\textwidth,height=6cm,draft=false]{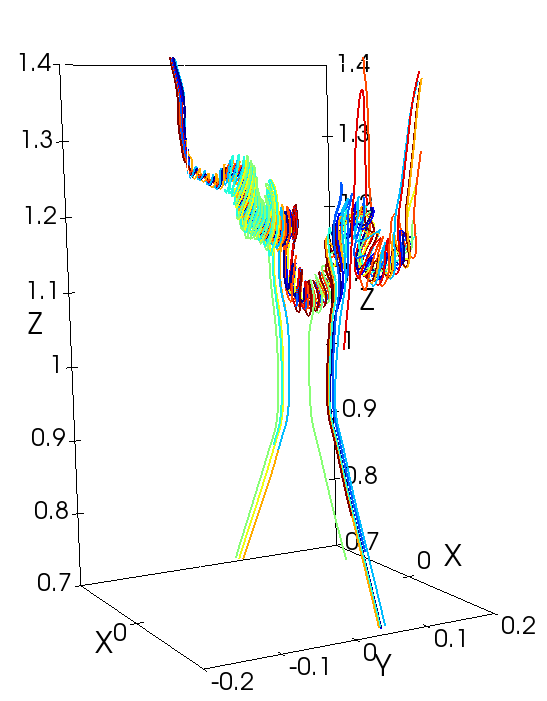}}
\subfigure[$t=1.42$]{\includegraphics[width=0.22\textwidth,height=6cm,draft=false]{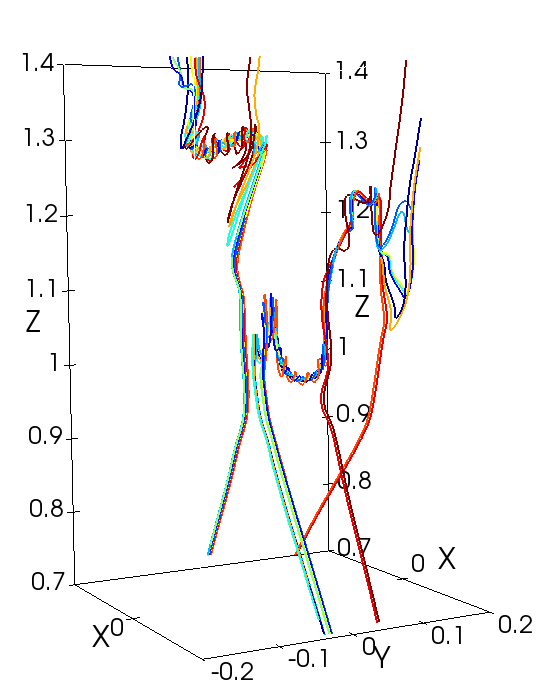}}
\subfigure[$t=1.44$]{\includegraphics[width=0.22\textwidth,height=6cm,draft=false]{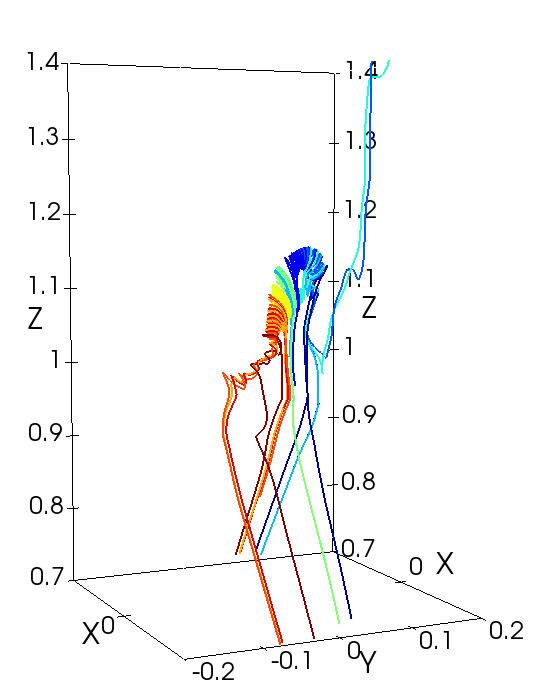}}
\caption{\small  Detailed evolution of the downward blobs from $t=1.38$ to $t=1.44$. (a)-(d) Structures of the blobs shown by the current isosurface with value $|J|=50$ colored by the velocity along the z-direction $V_z$ at the successive times. (e)-(h) 3D magnetic structures of the blob. Note that the colors for the magnetic field lines are used to better display the different lines. \label{down} }
\end{figure}

\cite{Lazarian2020} suggest that turbulent reconnection process dominates in 3D CS layers. However, the reconnection mechanisms proceeding in a CME-driven CS remain unclear. We follow the evolution of the downward blobs for four successive times just after the generation shown by Figure \ref{down}. In panel (a)-(d), the tube-like blobs marked by the black box are generated and move downward to interact with the flare loop-top. At $t=1.38$ in panel (e), two separate plasmoids are generated and recognized by unenclosed and highly twisted magnetic field lines. By plotting the field lines in different colors, we can observe the magnetic structures of blobs more clearly. Then at $t=1.40$ in panel (f), they begin to merge into a bigger plasmoid by changing the connectivity of the field lines, i.e. magnetic reconnection. This process of plasmoid coalescence is totally different from that in 2D, which yields a transverse (horizontal) CS of opposite current density between the mergers \citep{Barta2011}. The kinking of the plasmoid can also happen in the middle by obtaining extra twist turns. At $t=1.42$, the big plasmoid starts to collide with the flare loop-top, and is split into two separate plasmoids. The lower plasmoid loses its twist consecutively at one end during the collision, and the annihilation process of the plasmoid is likely to be related to turbulent reconnection occurring at the loop-top region. Lastly at $t=1.44$, the rest of the lower plasmoid also falls to the flare loops with two ends attached to the loop-top. The 3D complex structures and processes of plasmoids naturally provide multiple reconnection sites when merging into the flare loops.  \par
\begin{figure}[H]
\centering
\subfigure[$t=1.56$]{\includegraphics[width=0.26\textwidth,height=6cm,draft=false]{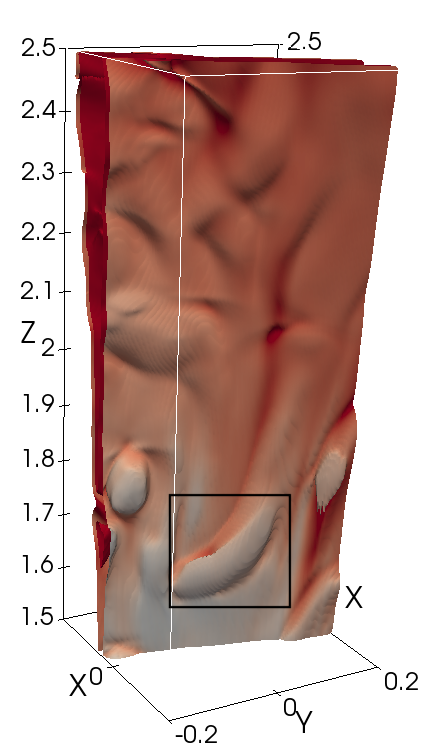}}
\subfigure[$t=1.60$]{\includegraphics[width=0.26\textwidth,height=6cm,draft=false]{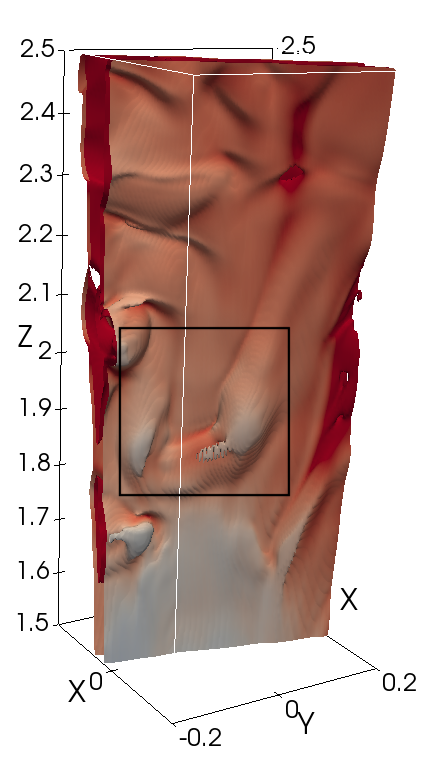}}
\subfigure[$t=1.64$]{\includegraphics[width=0.32\textwidth,height=6cm,draft=false]{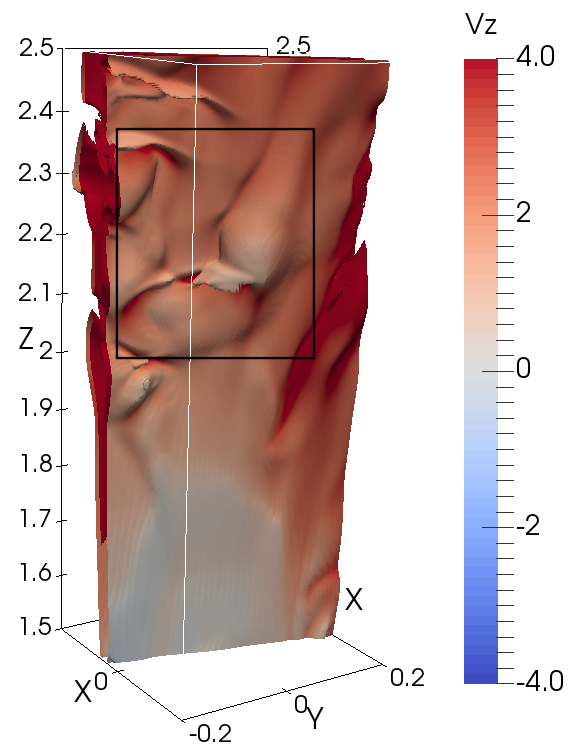}}
\subfigure[$t=1.56$]{\includegraphics[width=0.26\textwidth,height=6cm,draft=false]{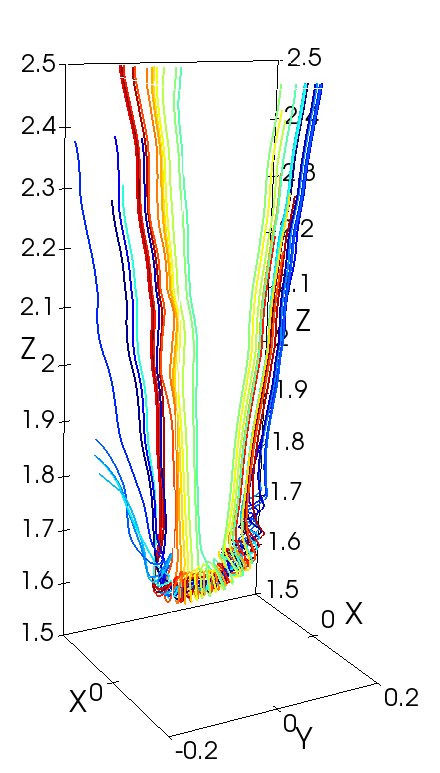}}
\subfigure[$t=1.60$]{\includegraphics[width=0.26\textwidth,height=6cm,draft=false]{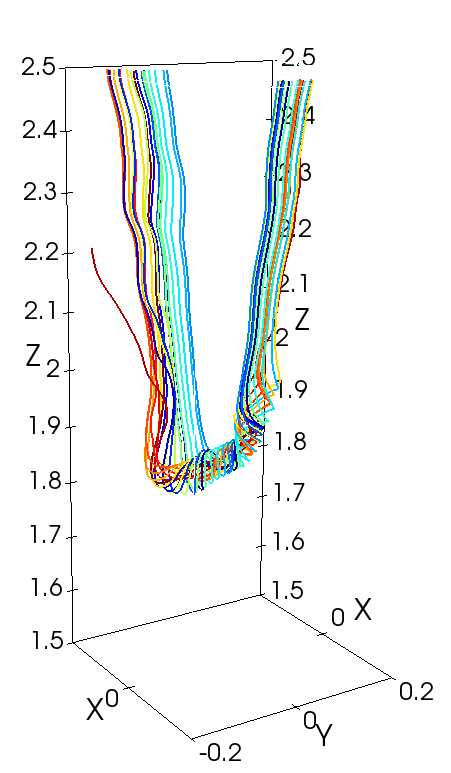}}
\subfigure[$t=1.64$]{\includegraphics[width=0.26\textwidth,height=6cm,draft=false]{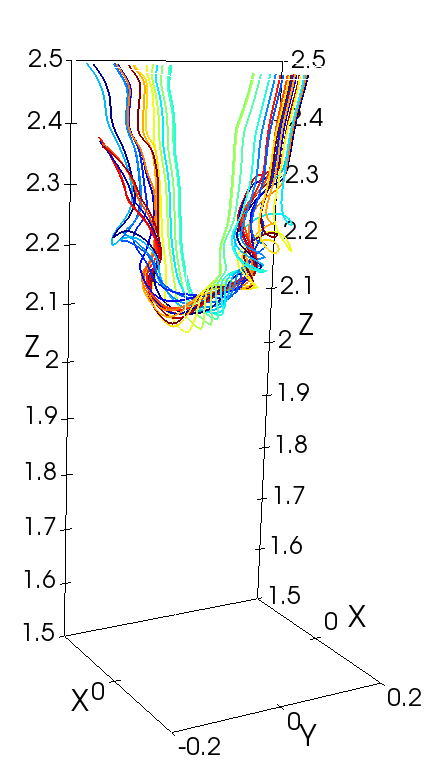}}
\caption{\small  Detailed evolution of an anti-sunward blob from $t=1.56$ to $t=1.64$. The physical terms are the same as Figure \ref{down}. \label{up} }
\end{figure}

On the other hand, we show the evolution of a blob moving upwards from $t=1.56$ to $t=1.64$ in Figure \ref{up}. In panel (a)-(c), a blob of 'U' shape leaves the PX-line with a speed of about $1285$ km/s and becomes eventually messed up far away from the PX-line in a short time. Regarding the field lines in panel (d), the blob is highly twisted in the center at $t=1.56$, which can be identified as a plasmoid. Then at $t=1.60$ in panel (e), the plasmoid swells while rising due to the low gas pressure at higher altitudes, and the turns of the flux tube are reduced. Later at $t=1.64$ in panel (f), the plasmoid only has  2 turns left at its two sides, and evolves gradually into a set of magnetic lines in a 'W' shape accompanied by SSs in various directions. This is to say, upward plasmoids are generated due to tearing instabilities, and evolve into the stochastic meandering of magnetic field lines (i.e. turbulence) in a short time, which induces a strong violation of magnetic flux freezing and a rapid energy dissipation \citep{Eyink2013}. Hence, the blob-like structures often observed in the flare CS \citep{Lin2005,Cheng2018} are not necessarily plasmoids, but instead turbulent post-plasmoid structures, because they were either well-developed plasmoids or merging/tearing magnetic islands. Their properties could be much different from other turbulence (e.g., around the PX region). Some of the blob-like structures seen in the flare CS have much different ionization states than the rest of the CS, for instance the ones seen in C III and O VI lines in the work of \cite{Ciaravella2008}. Those probably are separate plasmoids. However, the ones seen in white light could be either plasmoids or turbulent features.

 As discussed by \cite{Jiang2021}, the turbulent structures of CS layers become more complex in higher resolution computations. When the CME system evolves into a sufficiently turbulent stage, magnetic reconnection in the CS proceeds in a more complicated way rather than only Sweet-Parker or Petschek type. From our simulations, the downward plasmoids can keep the twisted shape until the annihilation at the flare loop-top, while many upward plasmoids turn into the turbulent post-plasmoid structures before arriving at the bottom of the FR. We can thus infer that turbulent reconnection dominates in the upper CS in 3D while plasmoid reconnection only dominates in the lower CS. 
\section{Conclusion and Discussion}\label{conclusion}
Plasmoid reconnection plays a governing role in the 2D magnetized plasma, and the dissipation scale, spectral properties and energy conversion process in the large-scale CS are exhaustively studied by theoretical and numerical work \citep{Lin2007,Shen2011,Ye2019,Ye2020,Zhang2022}. In the realistic 3D plate-like CS formed in a solar eruption \citep{Lin2002b}, the CS is highly confined by the reconnection inflows, so is the turbulence inside. However, the evolution of the confined turbulence in the CME-flare CS remains rarely reported in 3D. Aiming on this, we have performed full 3D MHD simulations for a moderate CME based on the analytical solution of \cite{Titov1999} in the gravitationally stratified atmosphere. More details are revisited and reported in the 3D view as follows:
\begin{itemize}
\item[(1)] Two runs with and without thermal conduction successfully reproduce the erupted FR with the propagation speed over $300$ km/s. The magnetic energy of the initial system converts into the kinetic and thermal energies, between which the kinetic energy dominates in the entire simulation time.
\item[(2)]  A plane-like CS above the flare develops and deflects following the FR motion. Many blobs in different shapes are generated in the CS due to tearing and plasmoid instabilities, and they undergo the splitting, merging and kinking processes in a more complex way than 2D cases. The plasma in the CS is mainly heated by the adiabatic and numerical viscous terms, and thermal conduction is the dominant factor that balances the energy inside the CS. Accordingly, the temperature of the CS reaches about 20 MK, and the range of temperatures is relatively narrow. 
\item[(3)] The reconnection rate decreases from a value larger than 0.1 in the impulsive phase, then oscillates and decays at the same time, which is consistent with the 2011 December 25 CME observed by \cite{Song2018}. And the generation of plasmoids in the CS gives rise to a higher reconnection rate in local area and the local rates along the PX-line are enhanced by turbulence with different amplitudes depending on the location of plasmoids.
\item[(4)] The reconnection outflows are highly nonuniform in the 3D CS as turbulence develops, which strongly disrupt the height of the PX-line. From the face-on view, the finger-like structures of SADs can be seen above the post-flare loops in AIA 131$\mathring{A}$, which confirm the main results of \cite{Shen2022} in a more self-consistent way. Moreover, the blobs are slightly brighter than the background CS fan in AIA 131$\mathring{A}$, which might extend the spikes of SADs to a higher place above the loop-top. It means that the moving blobs can also be a part of SADs.
\item[(5)] To analyze spatially the turbulent features of the CS, the 3D Fourier transform is performed on the velocity fields at $t=1.42$. The width and elongation of the CS might vary with height, showing the complex geometry of the 3D CS and turbulence. The spectra surprisingly have a significant variation either at different heights of the CS or in different directions, given the limited inertial range. 
\item[(6)] The PX-line is located always close to the flare loops and separates the CS into two parts, causing the energy partition to be asymmetric. The annihilation processes of plasmoids are different in two bi-directional reconnection outflows. Indeed, the upward plasmoids turn into stochastic magnetic field lines in a short time after the generation, while the downward plasmoids keep the twisted field lines until their disappearance at the flare loop-top. So the sunward blob-like structures are definitely plasmoids, while the anti-sunward ones often found in observations \citep{Lin2005,Cheng2018} are not necessarily identified as plasmoids, but instead turbulent post-plasmoid structures. That is to say, as the CS develops long enough, the plasmoid reconnection dominates in the lower CS, and the turbulent reconnection dominates in the upper CS. 
\end{itemize}
In the work of \cite{Kowal2020,Lazarian2020}, they reported that the plasmoid instability is only important for the initial stage of reconnection in presence of 3D homogeneous turbulence. But in our numerical work, various reconnection processes proceed simultaneously during the eruption process, in which plasmoid reconnection plays a key role for the lower CS. However, a precise calculation of the spectral slopes with respect to turbulence is still lacking due to the limited resolution in local place. A 3D high-resolution simulation of sufficiently large Lundquist number ($S=10^5-10^6$) for a solar flare will be performed in the future to determine the dissipation scale and CS thickness .

\acknowledgments
The authors would like to thank the anonymous referee for the valuable comments and suggestions to improve this work significantly. We are grateful to Chengcai Shen for fruitful discussions. This work is supported by National Key R\&D Program of China No.  2022YFF0503800, National Natural Science Foundation of China (NSFC) grants 12073073 and 11933009, grants associated with the Yunnan Revitalization Talent Support Program, the Foundation of the Chinese Academy of Sciences (Light of West China Program), the Yunling Scholar
Project of Yunnan Province and the Yunnan Province Scientist Workshop of Solar Physics, and grants 202101AT070018 and
2019FB005 associated with the Applied Basic Research of Yunnan Province. M.Z. acknowledges the support by the NSFC grants 12273107 and U2031141. C.Q. was supported by the NSFC grant 12203020 and the Yunnan Key Laboratory of Solar Physics and Space Science (202205AG070009). The calculations were carried out at the National Supercomputer
Center in Tianjin on the Tianhe 1A, and the numerical methods were tested on the cluster in the Computational Solar Physics Laboratory of Yunnan Observatories.

\appendix
\section*{Effect of Thermal Conduction}
This section is devoted to comparing the models without and with thermal conduction accomplished by our 3D MHD simulations, namely Run A and Run B listed in table \ref{para}. In Run B, the energy equation is solved with the parallel and perpendicular heat conduction formalized by 
\begin{equation}
\textbf{F}_C=-\kappa_{||}(\nabla T\cdot \hat{\textbf{B}})\hat{\textbf{B}}-\kappa_{\perp}(\nabla T-(\nabla T\cdot \hat{\textbf{B}})\hat{\textbf{B}}),
\end{equation}
where $\kappa_{||}=8\times 10^{-7}\frac{T_0^{7/2}}{\rho_0L_0v_A^3}T^{5/2}$ and $\kappa_{\perp}=4\times 10^{-10}\frac{n_0^2}{B_0^2T_0^3}\frac{\rho^2}{B^2T^3}\kappa_{||}$. The heat flux can saturate when temperature gradients become extremely steep, and is limited by the saturated flux suggested by \cite{Cowie1977}: 
\begin{equation}
\textbf{F}_{C,\textrm{sat}}=5\phi\rho c_{\textrm{iso}}^3,
\end{equation}
where $c_{\textrm{iso}}$ is the isothermal sound speed, and $\phi=1.1$ for the fully ionized gas.  \par
\begin{figure}[H]
\centering

\includegraphics[width=0.7\textwidth,height=15cm,draft=false]{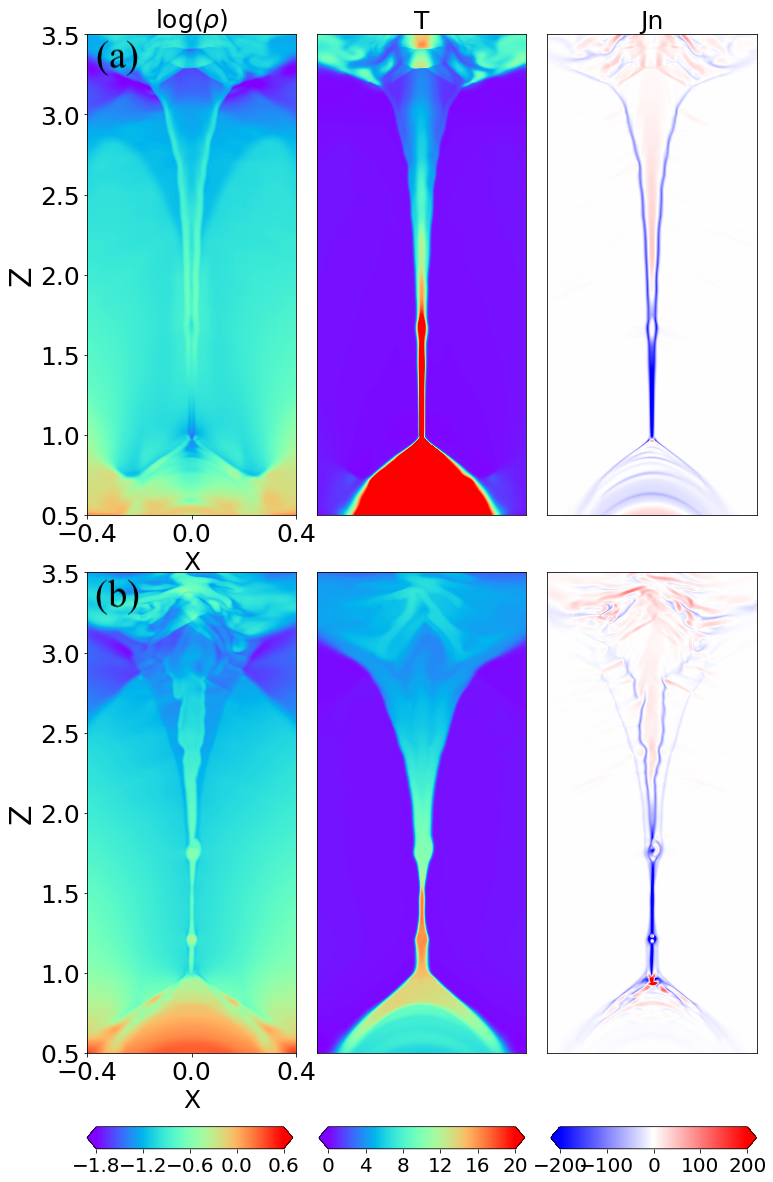}

\caption{\small Log of density (first column), temperature (second column) and current normal to plane for the $xz$-plane at the center ($y=0$) at $t=1.58$. (a) Run A; (b) Run B. \label{appendix} }
\end{figure}

Figure \ref{appendix} plots the density, temperature and current normal to plane for the $xz$-plane at the center ($y=0$) at $t=1.58$ for different runs. At this time, the CSs for Run A and B have already developed thin and long enough to let the instabilities take place. Blobs can be seen from the current density for both Runs, in which Run B shows less symmetry than the other. The density inside the CS and blobs is even lower than the background for Run A, while the CS and blobs are denser places for Run B. The temperature along the CS ranges between 3.6 and 17 MK for Run B, and the super hot structures ($>10$ MK) are mainly in the region above the flare loops. For Run A, the maximum temperature reaches 200 MK along the CS, and all the flare loops have the temperature exceeding 20 MK. As a result, thermal conduction is important to balance the energy inside the CS and yields the more realistic density and temperature distributions to compare with the observations. Together with Figure \ref{heat2D} and \ref{heat1D}, thermal conduction is shown to be a dominating factor for the heat transport when the plasma is heated in the solar eruption. 

%

\vspace{5mm}


\software{MPI-AMRVAC}


\bibliography{sample63}{}
\bibliographystyle{aasjournal}



\end{document}